\documentclass[
 reprint,
 amsmath,amssymb,
 aps,
  twoside,
 superscriptaddress,
prl
]{revtex4-1}
 
\usepackage{physics}
\usepackage{dsfont} 
\usepackage{color,newfloat}
\usepackage{graphicx}
\usepackage{dcolumn}
\usepackage{bm}
\usepackage{amsmath,amssymb}

\usepackage{array}  %packages for tables
\usepackage{hhline}
\usepackage{multirow}
\usepackage{afterpage}
\usepackage[normalem]{ulem}

\def\({\left(}
\def\){\right)}
\def\[{\left[}
\def\]{\right]}  

\DeclareMathOperator*{\argmax}{arg\,max}

\newcommand{\ttt}[1]{\text{#1}}

\begin{document}

\title{
Generation and sampling of
quantum states of light
in a silicon chip
}

\preprint{APS/123-QED}
\author{Stefano Paesani}
\altaffiliation{SP and YD contributed equally.}
\affiliation{Quantum Engineering Technology Labs, H. H. Wills Physics Laboratory and Department of Electrical and Electronic Engineering, University of Bristol, BS8 1FD, Bristol, United Kingdom}
\author{Yunhong Ding}
\email{yudin@fotonik.dtu.dk}
\altaffiliation{SP and YD contributed equally.}
\affiliation{Department of Photonics Engineering, Technical University of Denmark, 2800 Kgs. Lyngby, Denmark}
\affiliation{Center for Silicon Photonics for Optical Communication (SPOC), Technical University of Denmark, 2800 Kgs. Lyngby, Denmark}
\author{Raffaele Santagati}
\author{Levon Chakhmakhchyan}
\affiliation{Quantum Engineering Technology Labs, H. H. Wills Physics Laboratory and Department of Electrical and Electronic Engineering, University of Bristol, BS8 1FD, Bristol, United Kingdom}
\author{Caterina Vigliar}
\affiliation{Quantum Engineering Technology Labs, H. H. Wills Physics Laboratory and Department of Electrical and Electronic Engineering, University of Bristol, BS8 1FD, Bristol, United Kingdom}
\author{Karsten Rottwitt}
\author{Leif K. Oxenl{\o}we}
\affiliation{Department of Photonics Engineering, Technical University of Denmark, 2800 Kgs. Lyngby, Denmark}
\affiliation{Center for Silicon Photonics for Optical Communication (SPOC), Technical University of Denmark, 2800 Kgs. Lyngby, Denmark}
\author{Jianwei Wang}
\email{jianwei.wang@pku.edu.cn}
\affiliation{Quantum Engineering Technology Labs, H. H. Wills Physics Laboratory and Department of Electrical and Electronic Engineering, University of Bristol, BS8 1FD, Bristol, United Kingdom}
\affiliation{State Key Laboratory for Mesoscopic Physics and Collaborative Innovation Center of Quantum Matter, School of Physics, Peking University, Beijing 100871, China}
\author{Mark G. Thompson}
\email{mark.thompson@bristol.ac.uk}
\affiliation{Quantum Engineering Technology Labs, H. H. Wills Physics Laboratory and Department of Electrical and Electronic Engineering, University of Bristol, BS8 1FD, Bristol, United Kingdom}
\author{Anthony Laing}
\email{anthony.laing@bristol.ac.uk} 
\affiliation{Quantum Engineering Technology Labs, H. H. Wills Physics Laboratory and Department of Electrical and Electronic Engineering, University of Bristol, BS8 1FD, Bristol, United Kingdom}
\date{\today}

\begin{abstract}
\noindent
Implementing large instances of quantum algorithms
requires the processing of many quantum information carriers
in a hardware platform that supports the integration of different components.
While established semiconductor fabrication processes
can integrate many photonic components,
the generation and algorithmic processing of many photons
has been a bottleneck in integrated photonics.
Here we report the on-chip generation and processing of quantum states
of light with up to eight photons
in quantum sampling algorithms.
Switching between different optical pumping regimes,
we implement the Scattershot, Gaussian and standard boson sampling protocols in the same silicon chip,
which integrates linear and nonlinear photonic circuitry.
We use these results to benchmark a quantum algorithm for calculating molecular vibronic spectra.
Our techniques can be readily scaled
for the
on-chip implementation of
specialised quantum algorithms
with tens of photons,
pointing the way to
efficiency advantages over conventional computers.
\end{abstract}

\maketitle

\begin{figure*}[t]
  \centering
  \includegraphics[
  trim=0 0 0 10,
  width=1 \textwidth]{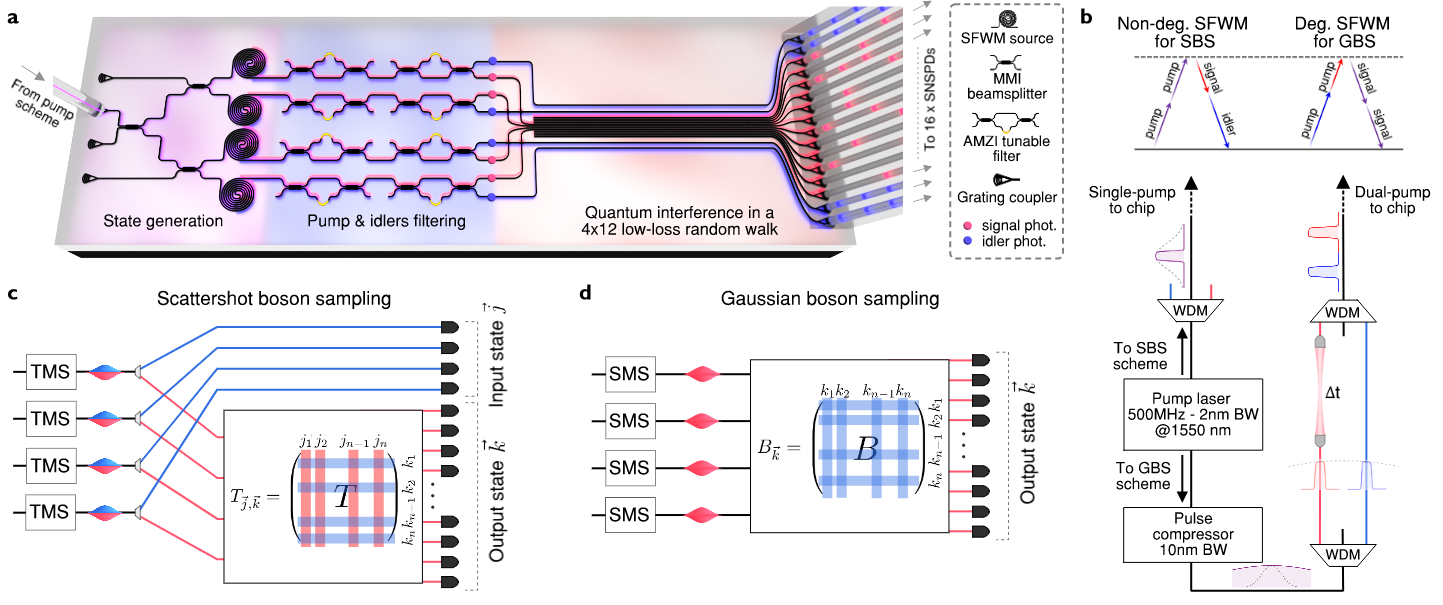}
  \caption{
Silicon photonic chip and experimental configuration.
(a)
The silicon chip integrates 
four SFWM spiral photon sources
and twelve continuously coupled waveguides
with a network of
MMIs and grating couplers;
AMZIs separate idler (blue)
and signal (red) photons,
and remove pump light (purple).
(b)
The SBS pumping scheme, where two-mode vacuum squeezing (TMS) is generated via non-degenerate SFWM, comprises
a $1550$ nm laser
with a bandwidth of $2$ nm,
and a WDM for wavelength selection.
In the GBS scheme, which relies on single-mode squeezing (SMS) generation via degenerate SFWM, 
a pulse compressor
increases the bandwidth to $10$ nm,
WDMs select dual wavelengths,
and a delay line synchronises the
arrival of the two pulses.
(c)
In a given run of the SBS protocol, 
the detection pattern
$\vec{j}$ measured in the idler modes
heralds the modes in which 
signal photons enter the random walk.
The probability to measure a given pattern
$\vec{k}$ after the random walk (described by a unitary matrix $T$)
is related to the permanent of
a sub-matrix of $T$,
whose rows and columns are defined by
$\vec{j}$ and $\vec{k}$ respectively.
(d)
In a given run of the GBS protocol,
four single-mode squeezed states are
generated and delivered to the random walk.
The probability for a given measurement pattern
in the GBS protocol is given by the Hafnian of sub-matrices of $B$, which is a function of both
$T$ and the squeezing parameters.
 } 
  \label{FigSetUp}
\end{figure*}

Devices that address customised problems
with quantum algorithms
are expected to demonstrate
an efficiency advantage over conventional computers.
Boson sampling is a specific model of quantum computing
that is suited to the platform of photonics
\cite{Aaronson2011,BroomeBS,SpringBS,CrespiBS,TillmannBS} 
and has been mapped to
the calculation of molecular vibronic spectra
\cite{AspuruGuzik},
simulation of spin Hamiltonians
\cite{Olivares2016},
simulation of molecular quantum dynamics
\cite{Sparrow2018},
and the enhancement of classical optimisation heuristics
\cite{XanaduGraphs}.
Implementing such applications at a size
that challenges conventional computers~\cite{Neville2017}
demands the integration and
high fidelity operation of
a large number of different components,
including
circuitry \cite{Carolan2015,Harris2017,16D},
detectors \cite{pernice2012},
filters \cite{Harris:2014kj},
and photon sources
\cite{Silverstone2013,Spring2017}.
The low efficiency of individual
spontaneous photon sources
has motivated the adoption of
deterministic solid-state photon sources \cite{Somaschi2016,Loredo2017,wang2017,BSloss}.
However,
the low-loss integration of solid-state sources
into photonic circuitry is an on-going challenge.

Creative approaches to realise boson sampling
with high numbers of photons
from spontaneous sources have seen
the design of variant models.
In principle,
the simultaneous optical pumping of
a number $k$ of spontaneous sources
that exceeds the number $n$ of desired photons,
boosts the overall rate of photon-pair production
combinatorially.
In Scattershot boson sampling (SBS)~\cite{Lund2014,Bentivegna2015,zhong2018Scattershot},
one photon from each pair heralds
the location of its partner,
such that a Fock state of $n$ photons is prepared
over a random subset of modes.
In Gaussian boson sampling (GBS)
\cite{GBS_Main},
$k$ single mode optical nonlinearities
are coherently pumped to produce $k$ modes of squeezed (vacuum) light, before linear optical processing and $n$-photon detection at the output.

The complex photonic circuitry required to scale these approaches can be addressed
with integrated photonics.
Here,
by pumping 4 integrated
spontaneous four wave mixing (SFWM) sources
with either a single-colour
or a two-colour laser,
we
select between
the Fock state required for SBS
and
the squeezed state required for GBS.
Both states of light are routed on-chip to
the same linear optical circuit,
which implements a random unitary operation over 
12 waveguides.
In the GBS mode of operation,
we benchmark this class of device
for calculation of
vibronic spectra.
In the limit of the SBS mode of operation,
with $n=k=4$,
we implement standard boson sampling
with 4 heralded photons,
generating and processing 8 photons
on-chip.
Our analysis shows
that larger versions
of our silicon photonic chip,
that exploit the combinatorial boost in photon rate available through the SBS and GBS protocols,
open up the regime
of
efficiency advantages over conventional computers.

The silicon circuitry and configuration of these experiments 
can be understood with reference to Fig.~\ref{FigSetUp}.
Four SFWM spiral sources are coherently pumped
by on-chip splitting of the
near-$1550\text{ nm}$ pump laser
via multi-mode interference (MMI) near 50:50 beam-splitters;
pump light is then removed
by asymmetric Mach-Zehnder interferometers (AMZIs).
In the dual-wavelength pumping (GBS) scheme
(where photons
are generated at the same signal wavelength)
two spectral regions of the
temporally compressed (spectrally broadened) pump
are selected and recombined using
wavelength-division multiplexers (WDMs),
before injection to the chip.
In the single-wavelength pumping (SBS) scheme,
(where signal and idler photons
are generated at different wavelengths)
idler photons are separated
using a second layer of AMZIs and
out-coupled to a fibre array to herald
the presence of the signal photons.
In both regimes, signal photons are routed to
the four central modes of
a continuous random walk,
implemented over 12 evanescently coupled waveguides,
then out-coupled to the fibre array.
Ultra low-loss out-coupling is implemented by
aluminium assisted apodized
grating couplers~\cite{Ding2014,Ding_Couplers}.
An array of 16 superconducting nano-wire single-photon detectors (SNSPDs) with approximately $78\%$ efficiency
detect the 4 heralding modes
and the 12 modes from the random walk.
(See Supplementary Materials for details).
%\\

\begin{figure*}[ht!]
  \centering
  \includegraphics[width=0.95 \textwidth]{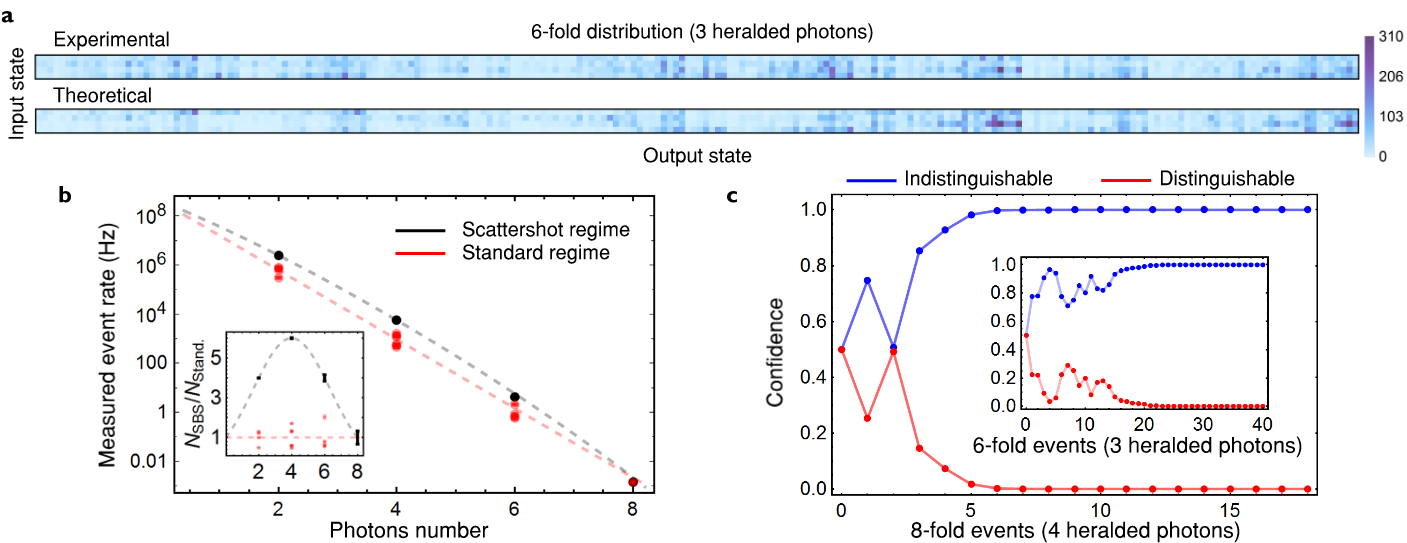}
  \caption{ \footnotesize	
Results for Scattershot boson sampling.
(a)
Experimental and theoretical distribution
for six-photon events,
with anti-bunched states
(one photon per detector) on horizontal axis
in increasing order $\{(1,2,3),(1,2,4),\ldots,(10,11,12)\}$ from left to right;
input states on the vertical axis are
$\{(1,2,3),(1,2,4),(1,3,4),(2,3,4)\}$
from bottom to top.
(b)
Measured event rates for 2 to 8 photons
are shown for the SBS (black points)
and standard (red points) regimes,
with a dashed line fit.
Inset is the photon rate enhancement
for SBS over the standard boson sampling protocol.
(c)
Dynamic Bayesian model updating
for validation that statistics are from indistinguishable rather than distinguishable photons for 8-photons (and 6-photons inset).
 } 
  \label{FigSSresults}
\end{figure*}

The SBS protocol is designed to tackle the
inefficiency that results from implementing standard boson sampling with spontaneous photon sources~\cite{Lund2014}.
In the original model of boson sampling~\cite{Aaronson2011},
a linear optical circuit is configured to implement
a random unitary operation over $m$ modes.
A number $n < m$ of input ports are each populated with a single photon, such that $n$ photons undergo an $n\times m$
random operation, before detection with photon counters.
Because the probability amplitude for each $n$-photon transition is equal to the permanent of the 
corresponding transfer matrix, which is in general intractable to classical computation,
an ideal experiment would efficiently produce samples from an essentially classically forbidden probability distribution.
However,
the rate at which the required $n$-photon state
is produced with spontaneous sources
decreases exponentially with $n$.
As shown in Fig.~\ref{FigSetUp}c,
the SBS protocol addresses this inefficiency 
by pumping $k$ sources that each produce weak two-mode squeezed light (with $ n < k \leq m$).
One (idler) photon from each pair is directly detected
to identify the input port of its partner (signal) photon. 
In comparison to the original scheme,
the probability to generate $n$-photons is boosted
by a factor $\propto {k\choose{n}}$ ~\cite{Lund2014,Bentivegna2015}.

We implemented the SBS protocol
in our silicon chip, with $k=4$
sources,
using the single wavelength pumping scheme
shown in Fig.~\ref{FigSetUp}b.
The experimentally measured distribution
for the $n=3$ case
(6-photon events)
shown in Fig.~\ref{FigSSresults}a,
has a mean fidelity \cite{Carolan2015}
with the theoretical distribution
of $92\%$.
Figure~\ref{FigSSresults}b shows
the measured difference
in photon pair generation rates
between the
SBS and standard boson sampling
(where only $n$ sources are pumped)
protocols for $n=1\text{ to }3$,
and for $n=4$ where the two regimes converge.
As predicted,
the $n$-photon pair generation rate,
is enhanced by a factor of
approximately $\binom{4}{n}$.
In the SBS regime,
the two-pair and three-pair photon rates
were measured at
$5.8\text{ kHz}$ and $4 \text{ Hz}$,
respectively,
while the 4-pair
(8-photon) rate
was approximately 4 events per hour.
The average purity of the signal photons,
estimated via
unheralded second-order correlation measurements
\cite{Christ2011},
was calculated at $86\%$.
(See Supplementary Materials for details).

While fidelity comparisons with theoretical distributions
are not a scalable method of validating boson sampling,
an efficient alternative is to compare measurements
against those predicted by a classically tractable
and plausible distribution.
As shown in Fig.~\ref{FigSSresults}c,
we used Bayesian model comparison
\cite{
bentivegna2014bayesian,
SpagnoloValidation,
Carolan2014,Bentivegna2015,Carolan2015}
to validate our $n=4$ (8-photon)
statistics against those predicted by distinguishable photons
(the $n=3$ case is inset).
The dynamically updated confidence that
samples are drawn from the distribution of
indistinguishable particles versus
distinguishable particles
reaches higher than 99.9\%.
(See Supplementary Materials
for details).

\begin{figure*}[ht!]
  \centering
  \includegraphics[width=0.95 \textwidth]{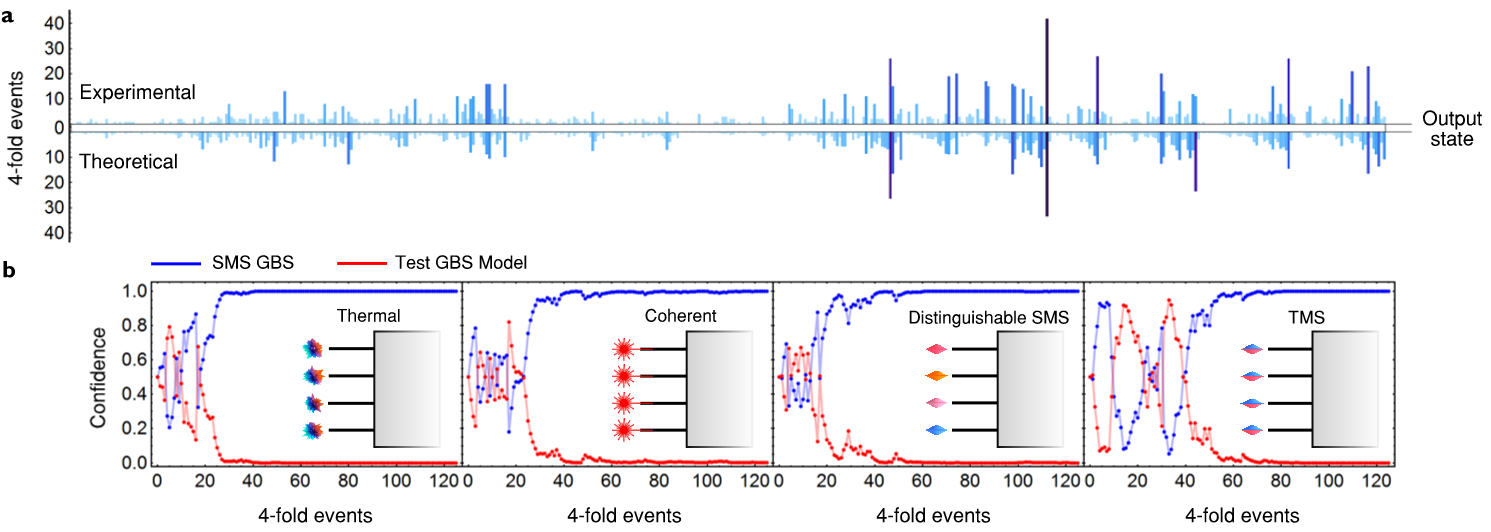}
  \caption{ \footnotesize	
Experimental results for GBS.
(a)
Experimental distribution for 4-photon events shows $87\%$ fidelity with the theoretical distribution.
The horizontal axis labels anti-bunched
(one photon per detector) states
in increasing order
$\{(1,2,3,4),(1,2,3,5),\ldots,(9,10,11,12)\}$
from left to right.
(b)
Results of dynamic Bayesian model updating
to validate that data is from the ideal GBS model
of single mode squeezed states (SMS GBS, blue lines)
against alternative (red lines) states,
from left to right:
thermal,
coherent,
distinguishable single-mode squeezed,
and two-mode squeezed states.
}  
  \label{FigGBS}
\end{figure*}

In contrast to SBS,
the GBS protocol does not project
the input state onto a single Fock state.
Rather, the input in GBS is an ensemble of single mode squeezed states, as shown in Fig.~\ref{FigSetUp}d,
which further increases the $n$-photon detection probability as compared to SBS~\cite{GBS_Main}.
After processing with linear optical circuitry
the probability for a particular pattern
at single photon detectors is given by a function
known as the Hafnian of the relevant transfer matrix.
Similar to the permanent of a matrix,
the Hafnian is computationally hard to calculate ~\cite{hafnian,GBS_Main,GBS_Details}.
(See Supplementary Materials for details).

We implemented the GBS protocol using
the two-colour pumping scheme
described in Fig.~\ref{FigSetUp}b,
which generates weak single-mode squeezed light at each source.
While this spectral selection 
reduces the pump power 
below that of the single-colour configuration,
we observed statistics with up to 4 signal photons at $1.1$ Hz rate. 
The experimentally measured distribution
for $n=4$ photons,
with one photon per detector
(full anti-bunching),
shown in Fig.~\ref{FigGBS}a,
has a mean statistical fidelity of $87\%$
with the ideal theoretical distribution.
In validating GBS,
a wide range of alternative models is available
where output distributions arising from general Gaussian input states are classically tractable.
As illustrated in Fig.~\ref{FigGBS}b,
we focus on models that are plausible in our
experiment.
We validate the ideal input state
of 4 single-mode squeezed states
against input with:
thermal states,
resulting from excessive loss;
coherent states, from unfiltered pump light;
distinguishable single-mode squeeze light,
due to spectral mismatch;
and two-mode squeezed states, 
from spurious photons generated at
different wavelengths.
For each test,
a confidence $>99.9\%$
is reached after $\approx 120$ samples
for an ideal model
using single mode squeezed states
rather than the alternative models.
(See Supplementary Materials for  details).

\begin{figure}[b]
  \centering
  \includegraphics[width=0.48\textwidth]{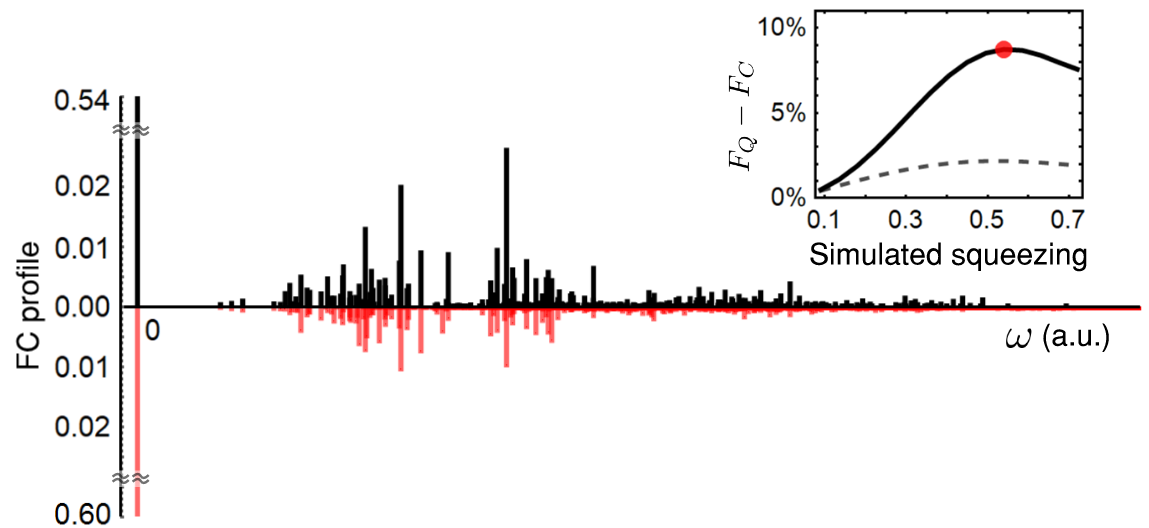}
  \caption{	
Reconstructed FC-profile.
Reconstructed FC-factors from
GBS data (black)
for frequencies $\omega$
are contrasted with
theoretical estimates (red)
for contributions of up to 8 photons
($>8$ photon contributions are negligible).
Inset is the improvement over optimal classical strategies
in the fidelity of estimating the FC-profile,
using simulated squeezing.
The solid line plots cases that simulate squeezing by processing experimental data,
while the dashed line indicates no post-processing.
The red point indicates the specific FC-profile plotted
in the main figure.}
  \label{FigFC}
\end{figure}

An application of GBS
is the calculation of
molecular vibronic spectra
\cite{AspuruGuzik,ClementsVibronic,shen2018},
where programmable linear optical circuitry,
together with squeezed and displaced light,
can approximate probabilities for the
vibrational transition (Frank-Condon factors)
between ground and excited states
of a given molecule.
To investigate the performance of
silicon photonic chips for this application,
we mapped our random walk circuit to a synthetic molecule
and considered the difference
$\mathcal{C}=F_Q - F_C$
between the fidelity of
the reconstructed Franck-Condon (FC) profile $F_Q$
and the optimal fidelity obtainable from
a classical strategy $F_C$ \cite{ClementsVibronic}.
While the FC-profile directly
reconstructed from GBS measurements 
has a fidelity $>99\%$ to the
ideal FC-profile,
the low level of squeezing
means that classically-tractable vacuum contributions dominate
and the improvement over a classical strategy
is small, ($\mathcal{C}=0.4\%$).

Data post-processing, based on characterised losses and detection efficiencies, makes it
possible to investigate 
how the fidelity of
the FC-profile
depends on the amount of simulated squeezing,
as shown in Fig.~\ref{FigFC}(inset).
For moderate levels of simulated squeezing,
an improvement of up to $\mathcal{C}=9\%$,
corresponding to $F_Q=86\%$ (shown in Fig.~\ref{FigFC}),
is obtained from post-processing.
For high values of simulated squeezing,
due to the increased contribution
of higher order photon terms,
the approach does not provide
any further advantage. 
(See Supplementary Materials for details).

Compared to SBS with 4 heralded photons,
we observed much higher rates for 4-photon GBS 
using a pump with lower power,
and requiring fewer detectors.
This type of resource saving for GBS over SBS
increases as the size of demonstration increases,
though GBS requires approximately twice as many photons as SBS
to demonstrate efficiency advantages over classical algorithms.
Increasing the number and
purity
of
integrated photon sources~\cite{caspani2017}
while decreasing the loss 
in photonic circuitry
to reduce noise and distinguishability among photons,
leads to similar fidelities when considering the scaling
of both the GBS and SBS protocols
(e.g.
loss acts to
add noise via thermal components
to the initial
SBS-two-mode
or GBS-single-mode
squeezed state resource
states).
The maximum number of photons we generated and processed
in these experiments is 8,
which is double the largest reported to date in integrated photonics.
Based on these results and further analyses
(see Supplementary Materials),
we estimate that arrays of several hundred
integrated detectors,
possible with current fabrication technologies~\cite{Schuck2013},
would allow fully-integrated experiments with tens of photons.
In the context of calculating molecular transition probabilities,
our class of photonic chip will be useful
for reconstructing Franck-Condon profiles
from photonic quantum sampling algorithms,
if relevant instances requiring $>20$ photon events are discovered.
More generally,
for problems that can be mapped 
to variant models of boson sampling,
such as interrogating the vibrational dynamics of molecules \cite{Sparrow2018},
our results and analysis show that
efficiency advantages over conventional computers
are a realistic prospect with the platform of integrated photonics.

\bibliography{biblio}

\begin{footnotesize}

\hfill \vspace{-0.1cm}

\noindent\textbf{Acknowledgements.} We acknowledge N. Maraviglia, R. Chadwick, C. Sparrow, L. Banchi, G. Sinclair and D. Bacco for useful discussions.  We thank W.A. Murray, M. Loutit, E. Johnston, H. Fedder, M. Schlagm\"uller, M. Borghi, and J. Lennon for technical assistance. We acknowledge support from the 
Engineering and Physical Sciences Research Council (EPSRC), European Research Council (ERC), and European Commission (EC) funded grants PICQUE, BBOI, QuChip, QITBOX, Quantera-eranet Square, and the Center of Excellence, Denmark SPOC (ref DNRF123).
Fellowship support from EPSRC is acknowledged by A.L. (EP/N003470/1).

\end{footnotesize}

%%%%%%%%%%%%%%%%%%%%%%%%%%%%%%%%%
%%%%%%%%    Appendix    %%%%%%%%%
%%%%%%%%%%%%%%%%%%%%%%%%%%%%%%%%%

\newpage 
\clearpage

\pagenumbering{arabic}

\onecolumngrid

\appendix

\setcounter{page}{1}

\newcommand{\hbAppendixPrefix}{S}
\renewcommand{\thefigure}{\hbAppendixPrefix\arabic{figure}}
\setcounter{figure}{0} 

\renewcommand{\thetable}{\hbAppendixPrefix\arabic{table}} 
\setcounter{table}{0}
\renewcommand{\theequation}{\hbAppendixPrefix\arabic{equation}} 
\setcounter{equation}{0}

%=============================================================================

\section{\centering\fontsize{15}{15}\selectfont 
\textbf{Supplementary Materials} }

\section{Experimental setup details}

\begin{figure*}[ht!]
  \centering
\includegraphics[width=1 \textwidth]{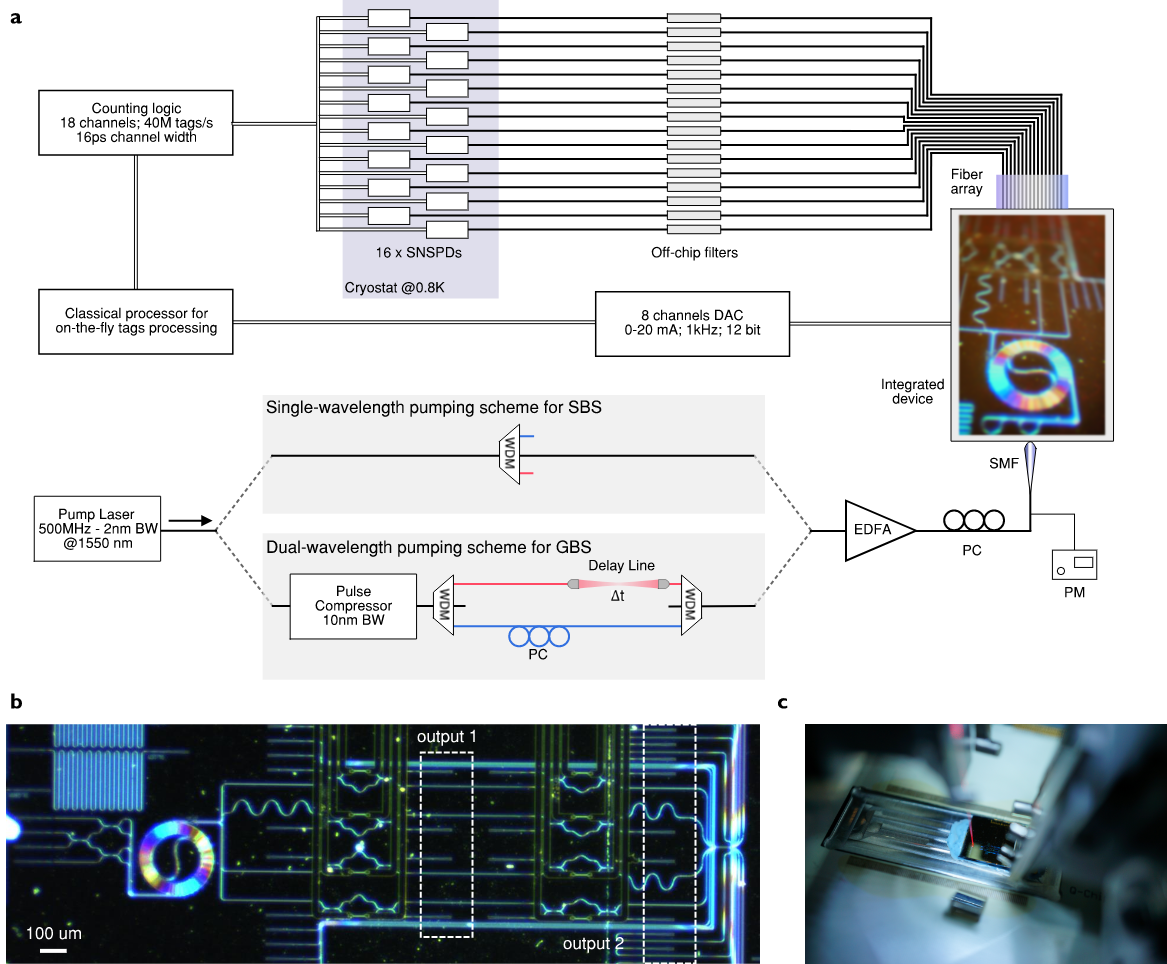}
  \caption{ \footnotesize	
Experimental setup schematic and images of the device. (a) Schematic of the experimental apparatus. WDM: wavelength-division multiplexer; PC: polarization controller; EDFA: Erbium-doped fiber amplifier; PM: photo-diode power meter; SMF: single-mode fiber; DAC: digital-to-analog converter; SNSPDs: superconducting nanowire single-photon detectors. (b) Optical microscope image of the integrated quantum photonic circuit. On the left the four sources, wrapped together in a spiral shape, can be observed, where photons are generated after the pump is coherently split by the MMI beam-splitters on the far left side. In the central region are the two layers of AMZI filters, with the electrical wiring used to control the thermal phase-shifters. On the right the continuous random walk, obtained by evanescently coupling 12 waveguides for a length of 110 $\mu$m, can be observed. The 12 outputs of the random walk and the 4 idler modes after the filters are routed to an array of 16 grating couplers (output 2), where photons are collected out of the chip via a fiber array (which was removed while taking this picture in order to show the photonic circuit below it). (c) Top view of the chip packaging, with the SMF on the left used for coupling pump light into the chip, and the fiber array on the right to collect the output photons. 
} 
\label{SIFig_ExpSchematic}
\end{figure*}

A detailed schematic of the experimental setup and images of the integrated device and the packaging are shown in Fig.~\ref{SIFig_ExpSchematic}.  Single photons were generated and guided in silicon waveguides with a cross-section of $450\text{ nm}\times 250\text{ nm}$, using spontaneous four-wave mixing (SFWM).  In the $\chi^{(3)}$ SFWM process weak squeezed states are generated as pump light is transmitted through four $1.4 \text{\ cm}$ long spirally-shaped waveguides~\cite{Silverstone2013}, see Fig.~\ref{FigSetUp}. In this process photons from the pump are annihilated while correlated pairs of photons are generated. Each of the 8 asymmetric Mach-Zehnder interferometer (AMZI) on-chip filters consist of two multi-mode interferometers (MMIs), acting as near-50\% beam splitters, and a thermal phase shifter on the shorter arm of the Mach-Zehnder interferometer. The interferometer is implemented via a continuous random walk obtained via evanescently coupling 12 waveguides, with the four sources injecting photons in the four central modes.  The distance between near-neighbour waveguides in the random walk is 450\,nm and the coupling region has a length of  $110 \mu\text{m}$, ensuring coupling between all 12 waveguides. In order to fine tune the AMZI filters, the chip was wire-bounded to a PCB to individually address each thermal phase shifter via an electronic voltage driver with 12 bits resolution (Qontrol\texttrademark).

\begin{figure*}[ht!]
  \centering
\includegraphics[width=1\textwidth]{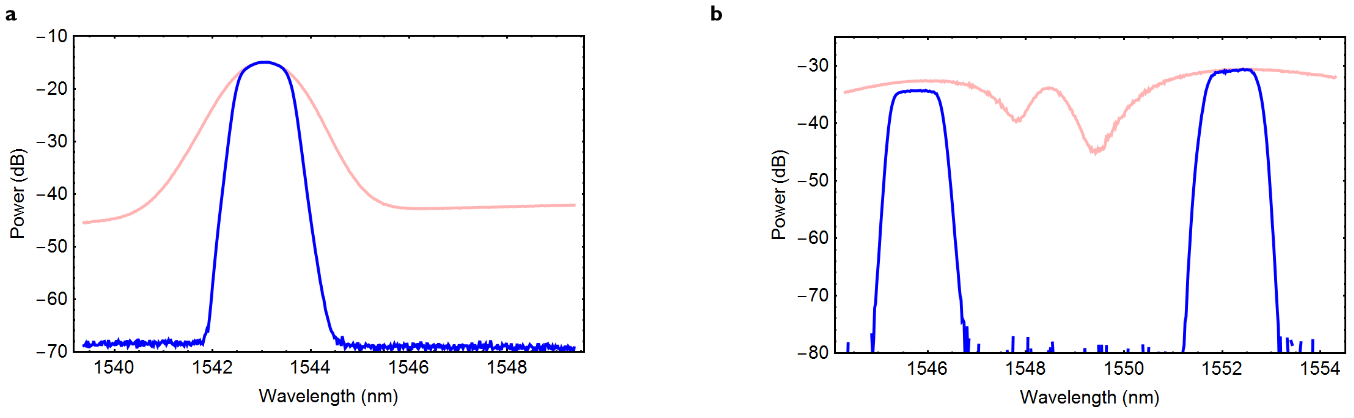}
  \caption{ \footnotesize	
Spectral traces of the input pump light for the two different pumping schemes. (a) Spectrum of the input pump light in the single-wavelength pumping scheme used for SBS. The red opaque line is the measured spectrum emitted by the laser, while the blue line is the spectrum of the light injected into the chip after passing through the WDM filter. (b) Spectrum of the input pump light in the dual-wavelength pumping scheme used for GBS. The red opaque line is the measured spectrum obtained after passing the laser pulses through the optical pulse compressor. The blue line is the spectrum of the light injected into the chip, obtained by selecting two spectral slices via a first WDM and recombining them into a single fiber using a second WDM.
} 
\label{SIFig_Spectra}
\end{figure*}

To pump the sources, laser pulses were generated via a tunable laser (PriTel\texttrademark Ultrafast Optical Clock) with a repetition rate of 500 MHz, emitting 2 nm bandwidth transform limited pulses at near-telecom wavelengths, and amplified via an Erbium-doped fiber amplifier (EDFA). In the single-wavelength pumping scheme, the emitted pulses (central wavelength of 1542.9 nm) are filtered off-chip via a wavelength-division multiplexer (WDM) ($>100$ dB extinction, 1.6 nm bandwidth) to remove spurious tails in the pump spectrum (see Fig.~\ref{SIFig_Spectra}a), and injected into the chip via a single mode fiber. Photons are emitted via non-degenerate SFWM at $1549.3$ nm (signal wavelength) and $1536.6$ nm (idler wavelength). In the dual-pump regime, the pulses, with a central wavelength of $1549.3$ nm are temporally compressed via a fiber optical pulse compressor (PriTel\texttrademark Femtopulse$^\text{\textregistered}$), achieving a 10 nm wide spectrum from which two slices (at 1552.5 nm and 1546.1 nm, each with 1.6 nm bandwidth) were selected via the WDM (see Fig.~\ref{SIFig_Spectra}b). A piezo-controlled optical delay line is used to ensure temporal overlap between the two obtained pulses of different wavelengths, which are then multiplexed together into a single mode fiber via a second WDM and injected into the chip. Photons are emitted on-chip via degenerate SFWM at the signal wavelength (1549.3 nm). In both pumping schemes, fiber polarization controllers were used to maximise the coupling. The input power is monitored via the use of a $99:1$ fiber beam-splitter and a photo-diode just before injecting the pump light into the chip.

Photons are collected out of the chip via a 16-channel fiber array, filtered via off-chip single channel fiber filters ($>100$ dB extinction, 0.8 nm bandwidth, 0.3 dB insertion loss on average) to remove spurious pump photons and enhance the photon indistinguishability, and finally sent to an array of 16 SNSPDs (Photon Spot\texttrademark, $\sim 100$ Hz dark  counts, $\sim 80\%$ efficiency) for detection. Fast counting logic (Swabian Instruments\texttrademark GmbH), supporting more than 40 million events per second, was used to collect the single photon detections and process them on the fly.

\section{Chip design, fabrication, and components characterisation}  
Optical losses, in particular insertion loss of integrated optical components and coupling loss, play a critical role for on-chip experiments. On our device, in order to decrease the coupling loss, a fully-etched apodized grating coupler using a photonic crystal is designed~\cite{Ding_Couplers}, and aluminium (Al) mirrors are used below the grating couplers via a flip-bonding process~\cite{Ding_Couplers}. The fabrication process starts from a commercial silicon-on-insulator (SOI) wafer with top silicon thickness of 250\,nm and buried oxide layer of 3\,$\mu$m, and proceeds as follows. Firstly, 1.6\,$\mu$m thick SiO$_2$ is deposited by the plasma-enhanced chemical vapour deposition (PECVD) process on the SOI wafer.  After that, the Al mirror is deposited by electron-beam (ebeam) evaporator, and followed by another thin layer of SiO$_2$ deposition with thickness of 1\,$\mu$m.  Following that, the wafer is flip-bonded with Benzocyclobutene (BCB) bonding process to another silicon carrier wafer. The substrate and buried oxide (BOX) layers of the original SOI wafer are removed by fast dry-etching and buffered hydrofluoric (BHF) acid chemical-etching, successively, resulting in the final Al-introduced SOI wafer. After that, the silicon photonic circuit is fabricated by standard ebeam lithography (EBL) followed by Inductively Coupled Plasma (ICP) etching and ebeam resist stripping. In order to simplify the fabrication process, grating couplers are designed as fully-etched so that they can be fabricated with the rest fully-etched silicon nanowires and other photonic components simultaneously. After the photonic part is fabricated, 1.3\,$\mu$m thick SiO$_2$ is deposited by PECVD, followed by polishing process to planarize the surface with approximately 300\,nm sacrifice, resulting in a final top SiO$_2$-cladding layer of 1\,$\mu$m. The micro-heaters are patterned afterwards by another EBL process followed by 100\,nm titanium (Ti) deposition and liftoff process. The conducting wires and electrode pads are obtained by standard Ultraviolet (UV) lithography followed by Au/Ti deposition and lift-off process.\\

\begin{figure*}[]
  \centering
  \includegraphics[width=1 \textwidth]{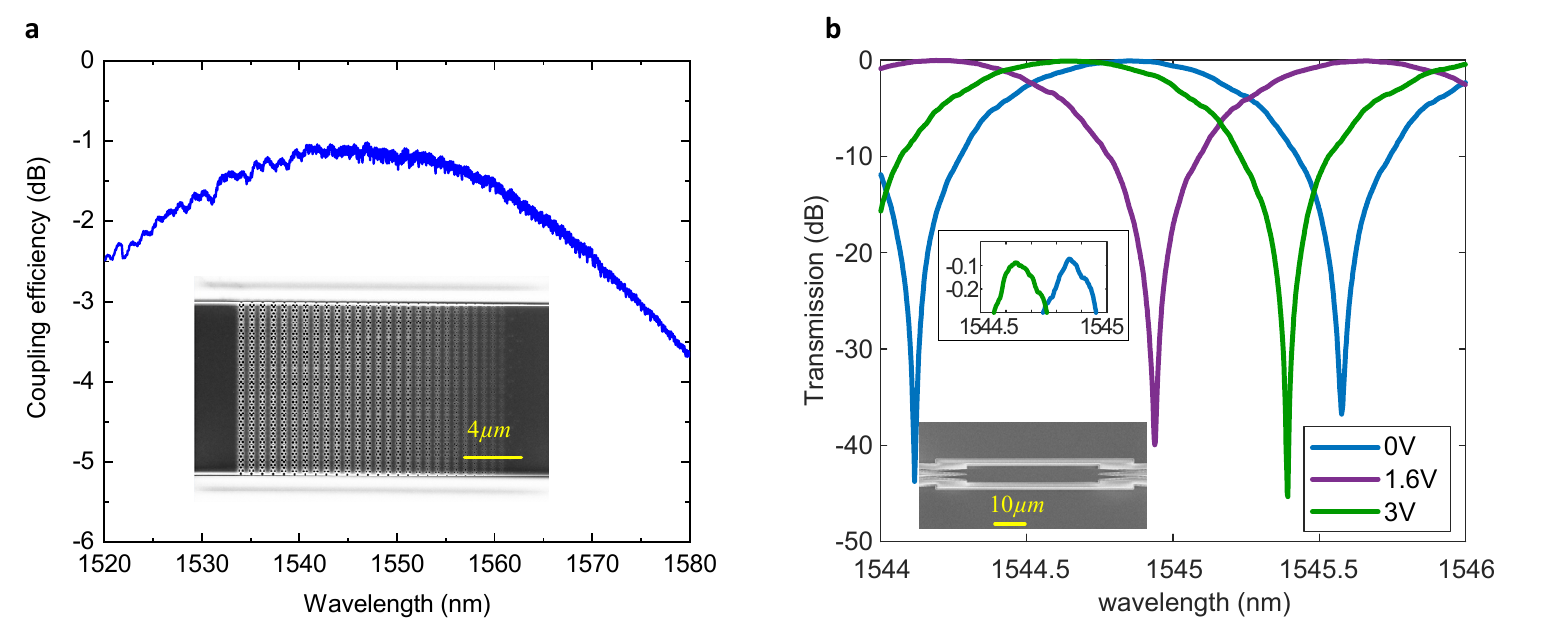}
  \caption{ \footnotesize	
Integrated components characterization. (a) Characterization of the low-loss grating coupler element. (b) AMZI characterization with different voltage applied on the heater. The scanning electron microscope (SEM) images of the components are also shown in the corresponding insets.
} 
\label{SIFig_components}
\end{figure*}

Our fabrication platform provide a propagation loss of $\sim$2\,dB/cm measured by the cut-back method for the fully-etched silicon waveguide with geometry of 450\,nm$\times$250\,nm. Figure~\ref{SIFig_components} presents the characterisations of the integrated components. The Al-mirror assisted grating couplers exhibit coupling efficiency of -1.1\,dB at the wavelengths used, with 1\,dB coupling bandwidth of 40\,nm, as shown in Fig.~\ref{SIFig_components}a. The 2$\times$2 MMI structures are used to implement AMZI filters with a titanium micro-heaters applied on one arm as phase shifters in the AMZIs tunable filters. In this situation, applying a voltage to the Ti micro-heater results in a power dissipated and heating the optical waveguide underneath, changing  the refractive index in the waveguide and inducing a phase shift. As shown in Fig.~\ref{SIFig_components}b, 3~\,V heating voltage results in a transmission shift of more than one free-spectral range (FSR). Moreover, the insertion loss presented in the inset of  Fig.~\ref{SIFig_components}b is less than 0.1\,dB, indicating less than 0.05\,dB insertion loss for each 2$\times$2 MMI.\\

\begin{figure*}[]
  \centering
  \includegraphics[width=0.9 \textwidth]{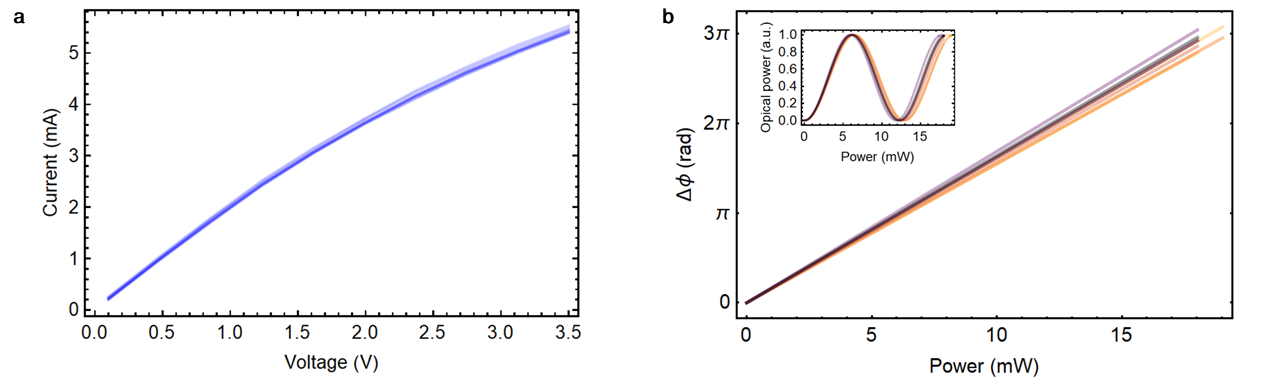}
  \caption{ \footnotesize	
Characterization of active components in the AMZI filters. (a) Current to voltage behaviour of the 8 thermal phase-shifters. (b) Characterisation of the optical phase implemented on each phase-shifter as a function of the thermal power dissipated. Inset: optical fringes in the AMZI filters used for fine tuning the filtered (destructive interference) and transmitted (constructive interference) wavelengths.
} 
  \label{SIFig_Heaters}
\end{figure*}

Inside the AMZI filters, the phase shifters allow to finely tune the filtered wavelengths. In order to achieve an accurate tuning, a precise characterisation of the heaters is required. The behaviour of the resistance for all heaters are presented in Fig.~\ref{SIFig_Heaters}a, indicating a consistent behaviour of all heaters. A non-Ohmic behaviour is observed, as is usual for Ti-based phase-shifters~\cite{Santagati2018,Paesani2017,16D}, and characterised in order to calculate the voltage required for a given heating power.  The phase-shift as a function of applied heating power is finally characterized for all the heaters, showing a consistent heating efficiency of all heaters with capability of $\sim$3$\pi$ for an applied heating power of approximately 18\,mW.\\

\begin{figure*}[ht!]
  \centering
  \includegraphics[width=0.8 \textwidth]{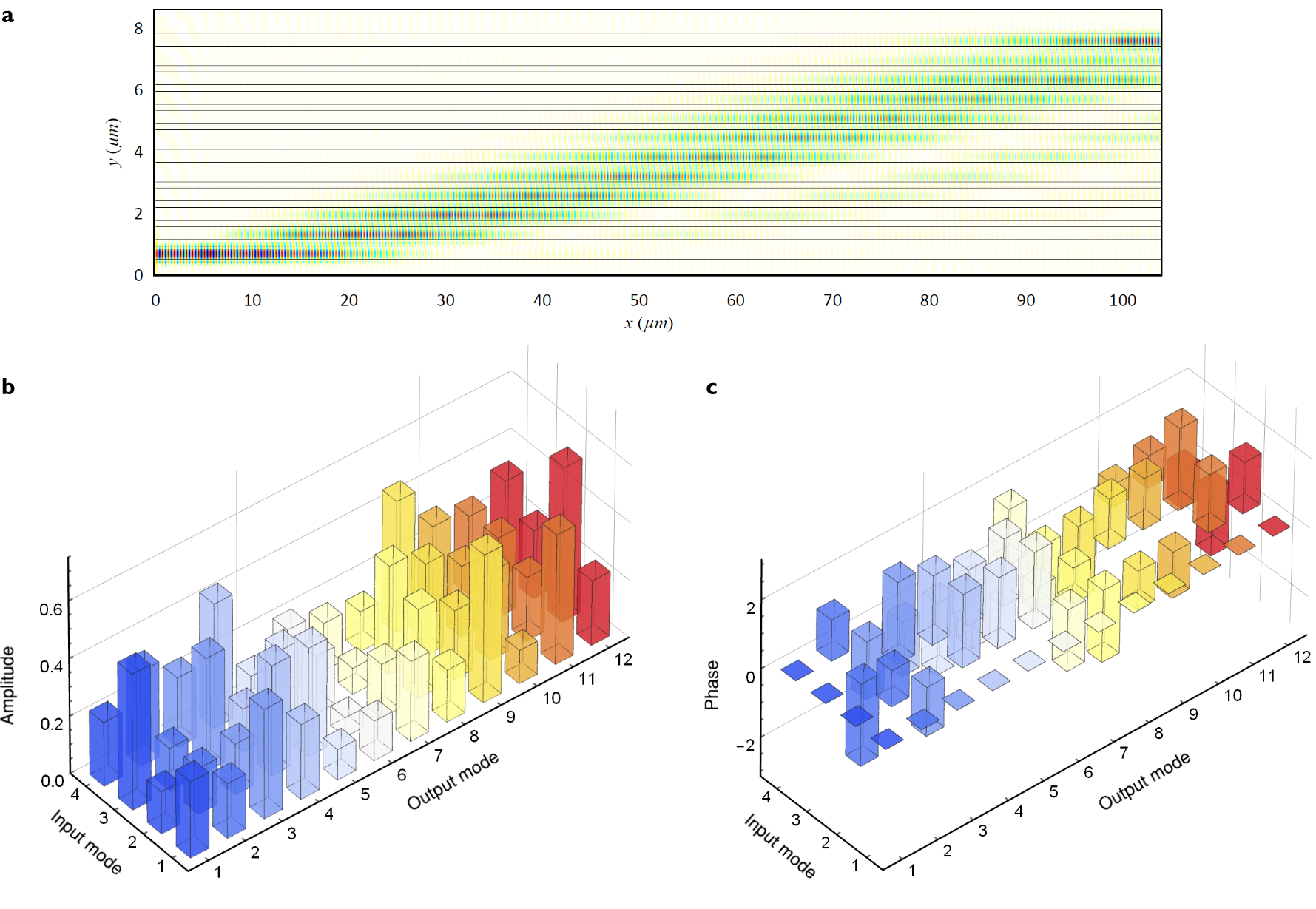}
  \caption{ \footnotesize	
Coupling FDTD simulation and characterisation of the random walk. (a) Simulation of 12$\times$12 coupled waveguides by 3D FDTD method for quantum Random Walk. In the simulation the waveguide design is the same as for the implemented circuit (waveguide width of 450~nm, coupling gap of 180~nm, coupling length of 110 ~$\mu$m) and assumes no phase fluctuations between the waveguides. It shows the evolution of light injected in the first waveguide to investigate the coupling between all waveguides. Amplitudes (b) and phases (c) of the measured entries of the $4\times 12$ transfer matrix.
} 
  \label{SIFig_RW}
\end{figure*}

In order to implement 4~$\times$12 random walk, a 12~$\times$12 waveguides evanescent coupler is designed. The widths of the 12 waveguides is designed to be 450\,nm with coupling gap of 180\,nm. The coupling length was chosen to be of $110 \mu\text{m}$ in order to ensure a coupling between all 12 waveguides, as was preliminarily determined by three-dimensional finite-difference time-domain (3D FDTD) simulations of the circuit, as reported in Fig.~\ref{SIFig_RW}a. The 3D FDTD simulations had only the scope to establish the length of the coupling region for the design, and were thus performed considering no phase fluctuations between the modes. In practice, random phase fluctuations naturally arise in silicon circuits due to nano-scale fabrication imperfections, which insert randomness in the unitary evolution of the random walk. Due to such effects it has been shown that the distribution of the obtained transform matrices converges to the Haar random distribution exponentially fast with the interaction length~\cite{Banchi2017}, making random walks a promising approach for the implementation of low-loss-interferometers for boson sampling~\cite{RandomWalksBS}. The transfer matrix of the implemented $4\times 12$ continuous random walk, which is the rectangular $4\times 12$ submatrix given by the 4 central rows of the total $12\times 12$ unitary matrix, is shown in Fig.~\ref{SIFig_RW}b-c, which was preliminarily characterised using standard  methods based on classical and two-photon interference~\cite{Laing2012,Peruzzo2011,Carolan2014,Carolan2015}. 

\section{System efficiency} 

Optical losses in the experimental set-up and in the integrated device were preliminarily measured. The average channel efficiency, that is the total loss experienced by each photon from generation to detection, measured via the coincidence to singles ratio~\cite{Silverstone2013} and averaged over all 16 channels used, was measured to be $-11.5$ dB. Characterisation of the 16 SNSPDs presented an average efficiency of $0.78$ (-1.0 dB). Mean transmission efficiencies through the off-chip filters and through the fiber connections to the detectors, averaged over the 16 channels used, were measured to be $0.87$ (-0.57 dB) and $0.94$ (-0.26 dB) respectively. 
Measured efficiency for the grating couplers is $0.77$ (-1.1 dB) at the wavelengths used, and the transmission for the AMZI on-chip filters is $0.98$ (-0.1 dB).  Propagation loss through straight waveguides was estimated via cut-back measurements to be approximately 2 dB $\text{cm}^{-1}$, indicating a transmission efficiency of $0.995$ for the 12-mode random walk implemented. Due to bending losses, the propagation loss in the spiral waveguides is higher, with average losses of $-7.1$ dB measured in the 1.4 cm long spirals.

\section{Degenerate and non-degenerate SFWM experimental details}

The performance of the individual sources was directly characterised by using the control grating couplers after the first array of AMZIs (outputs 1 in Fig~\ref{SIFig_ExpSchematic}b), configuring the filters accordingly to collect the photons before the interferometer.
Non-degenerate and degenerate SFWM effects were used for the generation of the input states required for SBS and GBS. The pumping schemes used to excite the processes are reported in Fig.~\ref{FigSetUp}b and Fig.~\ref{SIFig_ExpSchematic}a. For both regimes, the measured two-fold coincidences count rates from each single source are reported in Fig.~\ref{SIFig_SFWM}a for different input pump powers (the input pump power is measured off-chip, as shown in \ref{SIFig_ExpSchematic}a). To obtain these curves, for each source the two-fold coincidences are analysed using the same two channels and detectors, so that channel efficiencies are the same for all cases and differences in the curves are only due to the different efficiencies between the sources. The efficiencies of the two channels used, measured as $\eta_i=C_i/ CC$, where $C_i$ is the singles count rates on the associated channel and $CC$ are the coincidence counts, are $\eta_1=-9.7$~dB and $\eta_2=-8.9$~dB. 

In the non-degenerate SFWM regime (blue-green curves in Fig.~\ref{SIFig_SFWM}a), the coincidence count rates are given in term of the source efficiency (i.e. two-photon emission probability) $\epsilon=\tanh(\xi)^2/\cosh^2(\xi)$ as $CC=\epsilon R \eta_1 \eta_2$ (see for example Ref.~\cite{Silverstone2013,silverstone2015}), where $\xi$ is the squeezing parameter and $R$ is the repetition rate of the laser, from which $\epsilon$ can be calculated. In SBS measurement conditions (9 dBm input pump power) we obtain for the four non-degenerate SFWM sources 
the two-mode squeezing parameters $\vec{\xi}=\{0.25, 0.21, 0.18, 0.17\}$.

In the scheme used for the dual-pump regime, the pump is initially broadened in spectrum through a pulse compressor and the two different wavelengths are selected via WDM filters. This is done  to avoid synchronisation and phase locking issues that would arise when using two different pump lasers directly. While this significantly simplifies the experimental setup, it introduces some issues that have to be taken into account. First, a temporal delay between the two different frequencies is introduced when passing through the WDM filters, which is compensated via a delay line (see Fig.~\ref{FigSetUp}b and Fig.~\ref{SIFig_ExpSchematic}a). The photon count rates dependence on the temporal delay between the pump pulses with  different wavelengths is shown in Fig.~\ref{SIFig_SFWM}c. 
A peak of coincidences is observed when the pulses are overlapped, that is when photons are emitted via degenerate SFWM. A low level of noise  can observed when the two pumps are not overlapped, with a signal-to-noise ratio of approximately 30, which is mainly due to multi-pair emission of non-degenerate SFWM photons from the two different pumps. Secondly, by selecting only two frequencies on a broad pump spectrum, we significantly reduce the power injected into the chip, which in turn reduces the photon rates in the degenerate SFWM when compared to the non-degenerate SFWM case. In Fig.~\ref{SIFig_SFWM}a the photon pair rates (two-fold coincidences count rates) for each single degenerate SFWM source are reported (red-yellow curves) for different input pump powers. These were measured as in the non-degenerate SFWM case previously mentioned, with a only difference that a 50:50 off-chip fiber beam splitter was used to probabilistically separate the two degenerate photons (which succeeds with a one-half probability). Due to this probabilistic splitting, in the degenerate SFWM regime the coincidence count rates are given in term of the source efficiency $CC=\epsilon R \eta_1 \eta_2/2$ (again, see for example Ref.~\cite{Silverstone2013,silverstone2015}). As we are now generating weak single-mode squeezing, the source efficiency (i.e. two-photon emission probability) is given by $\epsilon=\tanh(\xi)^2/2\cosh(\xi)$. In the GBS measurement conditions (1 dBm total input pump power, measured off-chip) we obtain for the four degenerate SFWM sources  
the single-mode squeezing parameters $\vec{\xi}=\{0.11, 0.09, 0.07, 0.07\}$. These squeezing values have been used, together with the characterised unitary of the interferometer, to construct the matrix $B=T \lbrack \oplus_i \tanh(\xi_i) \rbrack T^t$ for GBS. In GBS measurement conditions (1 dBm input power), four-photon events are observed  at approximately a $5\text{ Hz}$ rate. 

While the count rate obtained using the implemented dual-wavelength pumping scheme for degenerate SFWM are greatly reduced by the low pump power injected, we remark that such scheme offers plenty of room for improvement. For example, the use of two separate pulsed lasers, each tuned at one of the two wavelength desired, would certainly allow much higher pump powers in the dual-wavelength regime, at the cost of requiring precise synchronisation between the two lasers.

\section{Photon purity characterisation}

\begin{figure*}[t!]
  \centering
  \includegraphics[width=1 \textwidth]{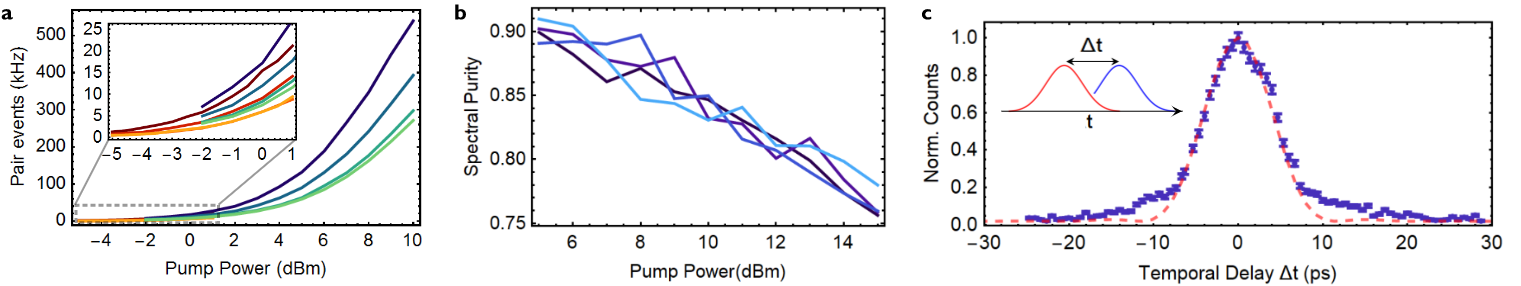}
  \caption{ \footnotesize	
Photon emission and purity via degenerate and non-degenerate SFWM. (a) Photon pair (two-fold) event rates measured at the detectors. Blue-green lines are for non-degenerate SFWM while red-yellow lines are for degenerate SFWM. In both cases, the source number and the darkness of the curve are associated: the darkest curve correspond to the first source (top mode in Fig.~\ref{FigSetUp}a) and the brightest to the fourth source (bottom mode in Fig.~\ref{FigSetUp}a). 
(b) Purity of the signal photons emitted with non-degenerate SFWM, obtained via unheralded $g_2$ measurements, for different input pump powers (measured off-chip). Curve darkness and source number are associated as in (a).  (c) Two-fold count rates emitted in the degenerate SFWM regime for different delays between the pump pulses with different wavelengths.  Error bars are obtained from Poissonian photon statistics.
} 
  \label{SIFig_SFWM}
\end{figure*}

In SBS high spectral purity of each signal photon emitted, heralded by the associated idler photon, is important to achieve a good quality quantum interference between the signal photons. While the good fidelities of the output distribution in the boson sampling experiments and the validation tests indicate good photon purities, in the non-degenerate SFWM case, where two-mode squeezed states are generated, a quantitative estimate of the photon purity can be obtained through unheralded second order correlation ($g_2$) measurements~\cite{Christ2011}. To perform these measurements, photons are collected at the output of each source (through the ports labelled as output1 in Fig.~\ref{SIFig_ExpSchematic}b), the idler and signal photons are separated through an off-chip filter and second order correlation measurements are performed on the signal photon while the idler is not measured. Following Ref.~\cite{Christ2011}, the photon purity $\mathcal{P}$ can be estimated from the obtained second order correlation $g_2(0)$ via $\mathcal{P}=g_2(0)-1$. In Fig.~\ref{SIFig_SFWM}b the obtained purities for the four different sources are reported for various values of the input pump power. A decrement of the purity for increasing pump powers is observed, plausibly due to non-linear noises in the sources and multi-pair emission errors, so that a trade-off between photon purity and generation rate has to be chosen. In measurement conditions for SBS via non-degenerate SFWM (9 dBm input pump power) the average purity is approximately $86\%$. 

\section{Bayesian model comparison for boson sampling validation.}

Boson sampling validation consists in providing supporting evidence to show that the measured samples are obtained from a true boson sampler, dictated by quantum interference, rather than from a classically tractable noisy device. 
In order to have a consistent verification protocol throughout the work, we have performed the validation via Bayesian model comparison techniques adapted from previous works~\cite{bentivegna2014bayesian,Carolan2014,SpagnoloValidation,Carolan2015}, whose generality allowed to validate all tested models using the same approach. The procedure is as follows. Suppose that, given a set of data $D=\{ x_i \}_{i=0}^N$ constisting of $N$ output samples $x_i$ from the interferometer, we want to verify if the data is more likely to arise from an ideal model $\bar{\mathcal{M}}$ or from a test model $\mathcal{M}_1$. For example, $\bar{\mathcal{M}}$ can represent an ideal implementation of SBS or GBS, while $\mathcal{M}_1$ a classical implementation where all photons are distinguishable. Given the measured set of samples $D$, Bayes' rule offers an immediate way to estimate the confidence in model $\bar{\mathcal{M}}$, that is probability for model $\bar{\mathcal{M}}$ to represent the underlying experiment, which is given by:
\begin{align*}
p(\bar{\mathcal{M}}|D)&=\frac{p(D|\bar{\mathcal{M}})p_0(\bar{\mathcal{M}})}{p(D|\bar{\mathcal{M}})p_0(\bar{\mathcal{M}})+p(D|\mathcal{M}_1)p_0(\mathcal{M}_1)} \\
&=\frac{1}{1+\frac{p(D|\mathcal{M}_1)p_0(\mathcal{M}_1)}{p(D|\bar{\mathcal{M}})p_0(\bar{\mathcal{M}})}}. 
\end{align*}
Assuming statistical independence between different events, the probability $p(D|\bar{\mathcal{M}})$ is given by $p(D|\bar{\mathcal{M}})=\prod_{i=1}^N p(x_i|\bar{\mathcal{M}})$, where $p(x_i|\bar{\mathcal{M}})$ is the probability of obtaining the measured outcome $x_i$ according to the ideal model (dictated by the permanent function for SBS and by the Hafnian in GBS), and similarly for $p(D|\mathcal{M}_1)$.

Note that, in the case where the detection is restricted to collision-free events, the probabilities $p(x_i|\bar{\mathcal{M}})$ have to be normalised to take it into account, which corresponds to dividing by the sum of the probabilities over all collision-free events. The reason for this is that the probability of a sample being collision-free differs between the different models; the normalisation constants are thus in general different.

The probabilities $p_0(\bar{\mathcal{M}})=p_0(\mathcal{M}_1)=0.5$ represent a prior distribution on the true model, which we here assume to be uniform to avoid any bias. 
The confidence $p(\mathcal{M}_1|D)$ for the adversary model $\mathcal{M}_1$ can be calculated in the same way, and both confidences are plotted in Fig.~\ref{FigSSresults}c for $\bar{\mathcal{M}}=\text{SBS}$ and in Fig.~\ref{FigGBS}b for $\bar{\mathcal{M}}=\text{GBS}$ as a function of the number of experimental samples, dynamically updating the confidence as new samples were collected. Note that, has the protocol requires to estimate probabilities $p(x_i|\bar{\mathcal{M}})$, which is not efficient for classical machines, the approach is not scalable. However, it is general and can  be used to validate against any model which allows to calculate probabilities $p(x_i|\mathcal{M}_1)$. 

One could also ask if it would be more convenient to obtain a simultaneous verification from a larger set of $m$ test models $\{\bar{\mathcal{M}},\mathcal{M}_1,\ldots,\mathcal{M}_m\}$, instead of performing multiple two-models  comparisons $\{\bar{\mathcal{M}},\mathcal{M}_k\}$. In such case, considering a uniform prior distribution between all models, Bayes' rule gives for the confidence in $\bar{\mathcal{M}}$
\begin{align*}
\frac{1}{1+ \frac{p(D|\mathcal{M}_{k_0})}{p(D|\bar{\mathcal{M}})}} \geq p(\bar{\mathcal{M}}|D)&=\frac{1}{1+\sum_{k=1}^m \frac{p(D|\mathcal{M}_k)}{p(D|\bar{\mathcal{M}})}}\\
&\geq \frac{1}{1+ m \frac{p(D|\mathcal{M}_{k_{\text{max}}})}{p(D|\bar{\mathcal{M}})}},
\label{BayesCompareMulti}
\end{align*}
with $k_{\text{max}}=\argmax_k p(D|\mathcal{M}_k)$ and $k_0 \in \{1,\ldots,m\}$ an arbitrary index value. From the equation above a link between the multiple comparisons with single models and the simultaneous comparison with multiple models can be observed. In particular, it can be easily verified that the confidence in $\bar{\mathcal{M}}$ in the simultaneous validation converges to one if and only if also the confidences in all the multiple two-model comparison validations  converge  to one.

\section{Scattershot boson sampling}

In the standard approach to boson sampling~\cite{Aaronson2011}, $n$-photon Fock states are generated by $n$ single photon sources and injected into the interferometer. However, if each source has a probability $\epsilon<1$ to generate a photon, then the probability that all $n$ sources fire is given by $\epsilon^n$, which decreases exponentially with $n$ and prevents any exponential quantum advantage. 
The SBS approach has been proposed to avoid these limitations and to increase the complexity of photonic experiments with realistic non-ideal photon-pair sources, such as those based on spontaneous parametric down-conversion or SFWM~\cite{Lund2014,Bentivegna2015}. As shown in Fig.~\ref{FigSetUp}c, in the SBS scenario, $k>n$ two-mode squeezing sources are simultaneously pumped, and for each source a possible detection of an idler photon on the idler mode heralds the presence of a signal photon in the associated mode. In this way a $n$ photon state is generated whenever a random, but heralded, set of exactly $n$ between the $k$ sources fire, which can happen with an enhanced probability $\binom{k}{n} \epsilon^n  (1-\epsilon)^{k}$, where now $\epsilon=\tanh(\xi)^2$ with $\xi$ the squeezing parameter. In the limit $k\gg n$, this probability represents an exponential combinatorial 
speed-up in the generation rate with respect to the standard approach, indicating that scalability can in principle be achieved even in presence of non-deterministic photon sources~\cite{Lund2014}. As the input state is an heralded $n$ photons Fock state, output probabilities are measured in the same way as for standard boson sampling, via the squared permanent of the submatrix associated to the input/output pattern~\cite{Aaronson2011,Lund2014}.

\section{Additional validation tests for scattershot boson sampling}

\begin{figure*}[t!]
  \centering
  \includegraphics[width=0.9 \textwidth]{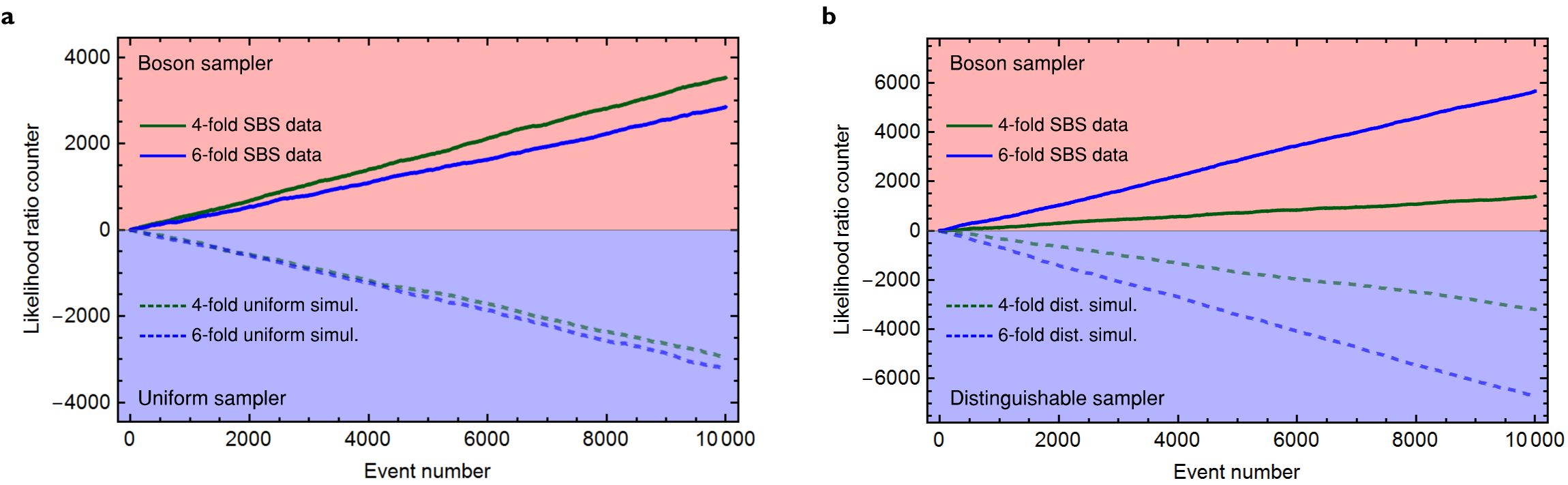}
  \caption{ \footnotesize	
Additional validations of SBS using the row-norm test and the likelihood ratio test. (a) Validation of the boson sampler against a uniform sampler via the row-norm estimator test by Aaronson and Arkhipov. (b) Validation of the boson sampler against a distinguishable sampler via the likelihood ratio test. Solid lines are obtained using the experimental SBS data, while dashed lines are obtained by running the test on simulated data generated from a uniform sampler (a) and a distinguishable sampler (b).
} 
  \label{SIFig_StandValid}
\end{figure*}

In Fig.~\ref{FigSSresults}c validation against a distinguishable sampler are reported for the SBS 6-fold and 8-fold data, using Bayesian model comparison techniques. The results give supporting evidence for the correct functioning of the experiment. To give further supporting evidence, we also performed validation protocols based on the likelihood ratio test and the row-norm estimator test, which are standards in previous experimental works~\cite{SpagnoloValidation,Bentivegna2015,wang2017,Loredo2017,BSloss}.

The row-norm estimator test, originally proposed by Aaronson and Arkiphov~\cite{Aaronson2014_sampling}, is designed to validate the outputs of a boson sampler from those obtained from a trivial uniform distribution. In contrast to the other validation protocols, this test can be efficiently performed (i.e. is scalable) as it does not require the calculation of ideal output probabilities. Considering a boson sampling experiment where $m$ output modes and $n$ photons are used, the row-norm estimator test proceeds as follows. The counter $C$ is initialised to $C=0$ and is iteratively updated each time an input/output sample $\{\vec{j},\vec{k}\}$ (here, $\vec{j}=\{j_1, \dots, j_m\}$ and $\vec{k}:=\{k_1, \dots, k_m\}$) is obtained from the experiment via calculating the row-norm estimator $R_{\vec{j},\vec{k}}:= \prod_p \sum_q |T_{\vec{j},\vec{k}} |^2 $, which is efficiently computable~\cite{Aaronson2014_sampling}, and using the updating rules
$$
C_{i+1}=\begin{cases} 
   C_i+1, & \text{if  } R_{\vec{j},\vec{k}} > (n/m)^m \\
   C_i-1, & \text{if  } R_{\vec{j},\vec{k}} \leq (n/m)^m.  
  \end{cases}
$$
After the data is collected, the protocol concludes that the
outcomes are drawn by a boson sampler if the final value of the counter is $C>0$, otherwise it concludes that the outcomes are drawn from a uniform distribution. The results of the row norm test using 4-fold and 6-fold data are shown in Fig.~\ref{SIFig_StandValid}a.

The likelihood ratio test is an alternative approach to distinguish an ideal boson sampler from a device which operates using distinguishable photons~\cite{SpagnoloValidation,Bentivegna2015}, and works as follows. For an ideal sampler and a distinguishable sampler the probability to observe an output state $\vec{k}$ given the input state $\vec{j}$ is given by $p_{\mathrm{ind}}(\vec{k}|\vec{j})=|\text{Perm}(T_{\vec{j},\vec{k}})|^2$ and $p_{\mathrm{dist}}(\vec{k}|\vec{j})=\text{Perm}(|T_{\vec{j},\vec{k}}|^2)$, respectively. In the likelihood ratio test, a counter $C$ is initialised to $C=0$ and is iteratively updated each time an input/output sample $\{\vec{j},\vec{k}\}$ is obtained from the experiment via calculating the estimator $L_{\vec{j},\vec{k}}=p_{\mathrm{ind}}(\vec{k}|\vec{j})/p_{\mathrm{dist}}(\vec{k}|\vec{j})$ and using the updating rules 
$$
C_{i+1}=\begin{cases} 
   C_i+2, & \text{if  } L_{\vec{j},\vec{k}} \geq a_2 \\
   C_i+1, & \text{if  } 1/a_1 \leq L_{\vec{j},\vec{k}} < a_2 \\
   C_i, & \text{if  }  a1 < L_{\vec{j},\vec{k}} < 1/a_1 \\   
   C_i-1, & \text{if  } 1/a_2 < L_{\vec{j},\vec{k}} \leq a_1 \\   
   C_i-2, & \text{if  } L_{\vec{j},\vec{k}} \leq 1/a_2.
  \end{cases}
$$
After the data is collected, the protocol concludes that the
outcomes are drawn by a boson sampler if the final value of the counter is $C>0$, otherwise it concludes that the outcomes are given by sampling distinguishable photons. In our tests, for the control parameters we used $a_1=0.75$ and $a_2=2$, with validation results from 4-fold and 6-fold experimental data reported in Fig.~\ref{SIFig_StandValid}b.

\section{Gaussian boson sampling}

The task of Gaussian boson sampling consists in generating samples from the photon-counting probability distribution of linear-optically evolved vacuum squeezed states. That is, we here deal with the probability $p_\ttt{gbs}(\vec{k})$ to detect a pattern of single photons $\vec{k}=\{k_1, \dots, k_m\}$ $(\sum_{i=1}^m k_i=n)$ at the output of a linear-optical circuit injected with $m$ vacuum squeezed states $\otimes_{i=1}^m\ket{\xi_i}$. In what follows, we shall characterize each state $\ket{\xi_i}$ in terms of its squeezing parameter $0\leq \xi_j<1$, while the linear-optical circuit in terms of the unitary matrix $U$ that transforms the input-mode operators $a_j^\dagger$ into output mode operators $b_k^\dagger$, $b_k^\dagger = \sum_j U_{kj} a^{\dag}_j$.

In the context of Gaussian boson sampling, it is useful to describe the linear-optically evolved state $\otimes_{i=1}^m\ket{\xi_i}$ via its Husimi Q-function~\cite{ulf}:
\begin{eqnarray}\label{a1}\nonumber
Q(\vec{\alpha})=\frac{1}{\pi ^m\sqrt{\, \ttt{det} \sigma_\ttt{Q}}}\exp\left[-\frac{1}{2}\vec{\pmb{\alpha}}^\dagger\sigma_\ttt{Q}^{-1}\vec{\pmb{\alpha}}\right],
\end{eqnarray}
where $\vec{\pmb{\alpha}}=(\alpha_1, \dots, \alpha_m, \alpha_1^*, \dots, \alpha_m^*)^\ttt{T}$, $\sigma_\ttt{Q}$ is the $2m\times 2m$ Q-covariance matrix of the evolved state,
\begin{eqnarray}
&&\sigma_\ttt{Q}=\frac{1}{2}\begin{bmatrix}
	{U} &  0  \\
	0         & {U}^*
\end{bmatrix}{S}{S}^\dagger\begin{bmatrix}
	{U}^\dagger &  0  \\
	0         & {U}^\ttt{T}
\end{bmatrix}+{I}_{2m}/2,\\
&&{S}=\begin{bmatrix}
	\oplus_{j=1}^m \cosh \xi_j   &      \oplus_{j=1}^m \sinh \xi_j   \\
	\oplus_{j=1}^m \sinh \xi_j    &      \oplus_{j=1}^m \cosh \xi_j
\end{bmatrix}
\end{eqnarray}
and ${I}_{2m}$ is the $2m\times 2m$ identity matrix. Next, using the formulas~\cite{ulf}
\begin{eqnarray}\label{a2}
&&p_\ttt{gbs}(\vec{k})=\pi^{m}\int_{\mathbb{C}^m}\prod_{i=1}^m d\alpha_i \, Q(\vec{\alpha})P_{\ket{\vec{k}}\bra{\vec{k}}}(\vec{\alpha}),\\ \label{a3}
&&P_{\ket{\vec{k}}\bra{\vec{k}}}(\vec{\alpha})=\prod_{i=1}^m e^{|\alpha_i|^2}\left(\frac{\partial^{2}}{\partial{\alpha_i}\partial{{\alpha_i}^*}}\right)^{k_i}\delta(\alpha_i)\delta(\alpha_i^*)
\end{eqnarray}
and integrating Eq.~(\ref{a2}) by parts using the generalized Fa\'{a} di Bruno's formula for partial derivatives~\cite{deriv}, we obtain the following expression~\cite{GBS_Main,GBS_Details},
\begin{eqnarray}\label{a4}
p_\ttt{gbs}(\vec{k})=\frac{|\ttt{Haf }B_{\vec{k}}|^2}{k_1!\cdots k_m!\sqrt{\ttt{det}\sigma_{\ttt{Q}}}}.
\end{eqnarray}
Here, $B_{\vec{k}}$ is a $n\times n$ matrix obtained from the $m \times m$ matrix $C={U}\cdot \ttt{diag}[\tanh\xi_1, \dots, \tanh \xi_m]\cdot{U}^\ttt{T}$ by repeating $k_i$ times its $i$th column and row. In turn, the Hafnian of a $2k\times 2k$ matrix $X$ is defined as~\cite{hafnian}
\begin{eqnarray}\label{a7}
\ttt{Haf}\,{X}=\sum_{\mu\in  \ttt{C}_{2K}} \prod_{j=1}^k {X}_{\mu(2j-1),\mu(2j)},
\end{eqnarray}
where C$_{2k}$ is the set of canonical permutations on $2k$ elements, obeying $\mu(2j-1)<\mu(2j)$ and $\mu(2j)<\mu[2(j+1)]$, $\forall j$. In fact, the Hafnian can be seen as a generalization of another matrix function, the permanent, which governs the statistics of indistinguishable bosons and is at the heart of the complexity of standard boson sampling models,
\begin{eqnarray}\label{a8}
\ttt{Perm}\,{X}=\sum_{\mu\in  \ttt{P}_{k}} \prod_{j=1}^k {X}_{1,\mu(j)},
\end{eqnarray}
where summation runs over all permutations of the numbers $\{1, 2, ..., k\}$. The computation of both matrix functions, permanents and Hafnians, is hard for a classical computer. Consequently, since the photon counting probability distribution $p_\ttt{gbs}(\vec{k})$ is given in terms of Hafnians, under certain complexity-theoretic assumption, the task of Gaussian boson sampling is intractable for a classical computer~\cite{GBS_Main,GBS_Details}.

Interestingly, scattershot boson sampling can be seen as a special case of Gaussian boson sampling. Namely, one can interfere pairs of adjacent vacuum squeezed states (input to the Gaussian boson sampler) on a beam splitter and use one of the arms of emerging two-mode squeezed states for heralding, while sending the other arms into a linear optical circuit $U$. This setting, being a special case of Gaussian boson sampling, results in a scattershot scenario~\cite{Levon_scattershot,Lund2014}. The corresponding photon counting probability distribution $p_\ttt{sbs}({\bf n})$ is now given in terms of permanents of the unitary matrix $U$,
\begin{eqnarray}
p_\ttt{sbs}(\vec{k})=\frac{|\ttt{Perm}\, U_{\vec{k},\vec{j}}|^2}{k_1!\cdots k_M!}.
\end{eqnarray}
Here, $U_{\vec{k},\vec{j}}$ is obtained from $U$ by repeating $k_i$ times its $i$th row and $j_i$ times its $i$th column, where the set $\vec{k}=\{k_1,\dots, k_m\}$ corresponds to the heralded pattern of single photons input to the circuit $U$.

We also remark that the linear-optical transformation that we deal with in our experiments has 12 output but 4 input modes and it is therefore given by a $4\times 12$ unitary transfer matrix $T$ (rather than a $12\times 12$ square matrix $U$). Such a scenario is equivalent to a standard setting with 12 input and output modes described by a $12\times 12$ square matrix $U$, just that all the eight extra input modes are injected with the vacuum. Effectively, this means that the corresponding rows of $U$ are deleted and we always deal with its $4\times 12$ submatrix $T$ obtained selecting the 4 central rows, which represent the transfer matrix of interest in our experiment.

\section{Test models for validating Gaussian boson sampling}

Bayesian approaches allow to validate experimental data by comparing an ideal experimental implementation of GBS against a general test model for which output probabilities are computable. In this section we describe in more details the test models used and how output probabilities can be computed and used in the validation tests.\\

{\bf Coherent states:} We first consider a set of $m$ input coherent states $\otimes_{i=1}^m \ket{\alpha_i}$, each of which has the following representation in the Fock basis, 
\begin{equation}\label{cs1}
\ket{\alpha_i}=e^{-\frac{|\alpha|^2}{2}}\sum_{n_i=0}^\infty\frac{\alpha^{n_i}}{\sqrt{n_i!}}\ket{n_i}.
\end{equation}

A linear optical circuit $U$ transforms a tensor product of coherent states $\otimes_{i=1}^m \ket{\alpha_i}$ into another tensor product of coherent states $\otimes_{i=1}^m \ket{\beta_i}$ with amplitudes
\begin{equation}\label{cs2}
\beta_k=\sum_{j=1}^M U_{kj} \alpha_j.
\end{equation}
In other words, coherent states remain separable while evolved through a linear optical circuit. Thus, the joint probability $p_{\mathrm{cs}}(\vec{k})$  of detecting a pattern of $n$ single photons $\vec{k}=\{k_1, \dots, k_m\}$, with an $m$-mode coherent state $\otimes_{i=1}^m \ket{\alpha_i}$ at the input, admits a simple product form
\begin{equation}\label{cs3}
p_\ttt{cs}(\vec{k})=\prod_{i=1}^{m}\frac{e^{-|\alpha_i|^{2}}|\alpha_i|^{2n_i}}{n_i!},
\end{equation}
which is a product of Poisson distributions. Consequently, both sampling from it and computing its elements can be done efficiently on a classical computer. Although this problem is trivial from a computational point of view, boson sampling from coherent states comprises a physically relevant test model for validating our GBS experiments. In the model implemented the input coherent states amplitudes were considered to be uniform between all modes, as is physically plausible.\\

{\bf Thermal states:} Our next test model deals with sampling from linear-optically evolved thermal states. That is, we consider an $M$-mode input thermal state $\otimes_{i=1}^m \rho_i^{\ttt{th}}$. In turn, each state $\rho_i^{\ttt{th}}$ can be expressed as an incoherent mixture of Fock states,
\begin{equation}\label{th1}
\rho_i^{\ttt{th}}=(1-\tau_i) \sum_{n_i=0}^\infty \tau_i^{n_i} \ket{n_i}\bra{n_i},
\end{equation}
where $\tau_i=\langle{n_i}\rangle(\langle{n_i}\rangle+1)$ and $\langle{n_i}\rangle$ is the average photon number of the state. The probability to detect a set $\vec{k}=\{k_1, \dots, k_m\}$ of $n$ single photons at the output of a linear-optical circuit $U$ injected with an $m$-mode thermal state $\otimes_{i=1}^m \rho_i^{\ttt{th}}$ then reads
 \begin{equation}\label{th2}
p_{\mathrm{th}}(\vec{k})=\frac{1}{k_1!\cdots k_m! }\frac{ \mathrm{Perm} \, {A}_{\vec{k}} }{\prod_{i=1}^m(1+\langle{n_i}\rangle)}   ,
\end{equation}
where
\begin{eqnarray}\label{th3}
&&A=U D U^\dagger, \\ \label{4.2}
&&{ D}=\ttt{diag}\left\{\tau_1, ..., \tau_m \right\}
\end{eqnarray}
and ${A}_{\vec{k}}$ is obtained from $A$ by repeating $k_i$ times its $i$th column and row.
 
Interestingly, although the photon-counting probability distribution for input thermal states is given in terms of matrix permanents (namely, of Hermitian positive semi-definite matrix permanents, since $\tau_i \geq 0$, $\forall i$) sampling from the probability distribution in Eq.~(\ref{th2}) is classically tractable~\cite{Ralph2015}. Even more, computing its elements can also be done efficiently for a restricted set of linear-optical transformations $U$ and input states $\rho_i^\ttt{th}$~\cite{Levon2017_perms}.\\

{\bf Distinguishable squeezed states:} The treatment of GBS with distinguishable squeezed states is analogous to that of standard boson sampling with distinguishable single photons. Namely, since distinguishable squeezed states do not interfere with each other, the corresponding GBS experiment with $k$ distinguishable squeezed states is equivalent to a set of $k$ experiments where a single squeezed state $\ket{\tilde{\xi}_j}:=\ket{0,\dots, 0, \xi_j, 0, \dots, 0}$  evolves according to the $m$-mode linear-optical transformation $U$ ($j=1, \dots, k$). In turn, the photon-counting statistics at its output is obtained by accumulating photon detection events from these experiments. To find the corresponding probability distribution, we first write down the Q-covariance matrix of the evolved states $\ket{\tilde{\xi}_i}$: 
\begin{eqnarray}\label{dss1}
\tilde{\sigma}^{(j)}_\ttt{Q}=\frac{1}{2}\begin{bmatrix}
	{U} &  0  \\
	0         & {U}^*
\end{bmatrix}\tilde{{S}}_j\tilde{{S}}_j^\dagger\begin{bmatrix}
	{U}^\dagger &  0  \\
	0         & {U}^\ttt{T}
\end{bmatrix}+{I}_{2m}/2
\end{eqnarray}
where 
\begin{eqnarray}\label{dss2}
\tilde{{S}}_j=\begin{bmatrix}
	\oplus_{i=1}^m (1+\delta_{j,i}\cosh \xi_j)   &      \oplus_{i=1}^m \delta_{j,i}\sinh \xi_j   \\
	\oplus_{i=1}^m \delta_{j,i} \sinh \xi_j   &      \oplus_{i=1}^m (1+\delta_{j,i}\cosh \xi_j) 
\end{bmatrix}.
\end{eqnarray}

Next, in order to find the probability $p_\ttt{dss}(\vec{k})$ of detecting a pattern $\vec{k}=\{k_1, \dots, k_m\}$ of $n$ single photons at the output of $k$ distinguishable squeezed states distributed among $m$ modes, one has to consider all the possible detection events upon the evolved states $\ket{\tilde{\xi}_j}$, $\forall j$, that yield the given pattern $\vec{k}$. Since the corresponding expression is rather bulky for arbitrary $n$, $k$ and $m$, we present here the expression relevant to our experimental setting only. That is, we assume that four distinguishable squeezed states are injected in the $a$th, $b$th, $c$th and $d$th mode of a $12$-mode linear-optical circuit. The probability $p_\ttt{dss}^{(4)}(\vec{k})$ of detecting $n=4$ single photons in the $q$th, $r$th, $s$th and $t$th modes (no more than two photons per mode) at its output then reads 
\begin{eqnarray}\label{dss3}\nonumber
p_\ttt{dss}^{(4)}(\vec{k})=&&\frac{1}{\prod_{f\in \{a,b,c,d\}}\sqrt{\det \tilde{\sigma}^{(f)}_{{\ttt{Q}}}}}\left(\sum_{\{i,j\}\in G}\left[\frac{p^{(i)}_{l_1,l_2}p^{(j)}_{l_3,l_4}}{(1+\delta_{l_1,l_2})(1+\delta_{l_3,l_4})}\right.+\frac{p^{(i)}_{l_1,l_3}p^{(j)}_{l_2,l_4}}{(1+\delta_{l_1,l_3})(1+\delta_{l_2,l_4})}+
\left.\frac{p^{(i)}_{l_1l_4}p^{(j)}_{l_2,l_3}}{(1+\delta_{l_1,l_4})(1+\delta_{l_2,l_3})}\right]\right.\\ 
&&\left.+\sum_{i\in \{a,b,c,d\}}\left|\ttt{Haf}\,\tilde{{C}}_i'\right|^2\right).
\end{eqnarray}
Here, $G$ is the set of all six distinct pairs of input mode numbers, i.e., $G=[\{a,b\}, \{a,c\}, \{a,d\}, \{b,c\}, \{b,d\}, \{c,d\}]$, $l$ is the list of mode numbers where photons were detected, $l=\{q, r, s, t\}$, $\tilde{{C}}_i'$ is obtained from the $12\times 12$ matrix $\tilde{C}_i={U}\cdot \ttt{diag}[0, \dots, \tanh\xi_i, \dots, 0]\cdot{U}^\ttt{T}$ by keeping its $q$th, $r$th, $s$th and $t$th rows and columns and $p^{(j)}_{ij}=\left|\tilde{{C}}^{(j)}_{ij}\right|^2$. If two photons were detected at a given mode, the corresponding mode number is repeated twice in the list $l$. For instance, if two photons were detected in the $r$th mode, the list $l$ reads $l=\{q, r, r, t\}$. Note also that the first line in Eq.~(\ref{dss3}) corresponds to events where a pair of squeezed states produced a detection of two photons, while the second line corresponds to events where a single squeezed state produced a detection of four photons.
 
We also remark that although the computation of each element of the photon-counting probability distribution $p_\ttt{dss}(\vec{k})$ is hard, sampling from it is classically tractable, analogous to the case of standard boson sampling with distinguishable single-photons (see, e.g., Ref.~\cite{Aaronson2014_sampling}).\\

\begin{figure*}[ht!]
  \centering
  \includegraphics[width=0.7 \textwidth]{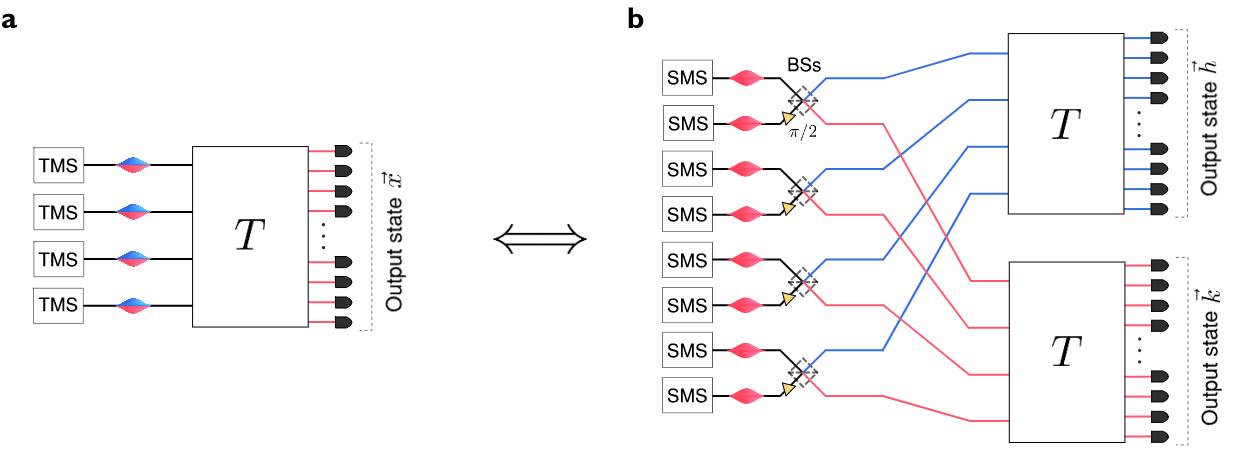}
  \caption{ \footnotesize	
Equivalence between a sampler with two-mode squeezed states at the input and a circuit with single-mode input squeezed states. (a) Photonic circuit in of a Gaussian boson sampler using TMS at the input. (b) Equivalent circuit with SMS at the input, where each TMS source is substituted with two SMS sources that are interfered into a beam-splitter after the bottom source has accumulated a $\pi/2$ phase.
} 
  \label{SIFig_TMSGBS}
\end{figure*}

{\bf Two-mode squeezed states:} To treat the case where the input states are two-mode squeezed states instead of single-mode ones, we can make use of the analogy between the photonic circuits represented in Fig.~\ref{SIFig_TMSGBS}a and Fig.~\ref{SIFig_TMSGBS}b. Similar connections between sampling protocols using TMS and SMS have already been highlighted in previous works, for example in Ref.~\cite{GBS_Details} it was used to relate SBS and GBS. The idea is to note that $m$ two-mode squeezers can be obtained from $2m$ single-mode squeezers combined pairwise into phase shifters and beam-splitters, as shown in Fig.~\ref{SIFig_TMSGBS}b. As photons in different modes of the two-mode squeezers would not interfere in the interferometer, we can separate all the $m$ top output modes of the beam-splitters and the $m$ bottom output modes, and send them into separate but equal interferometers ${T}$. Suppose that single-photon detection at the outputs of both interferometers, obtaining patterns $\vec{h}$ and $\vec{k}$ in the top and bottom $m$ output modes, respectively (note that $\sum_i h_i=\sum_i k_i=n/2$ must hold, where $n$ is the total number of photons). This would correspond to a detection patter $\vec{x}=\vec{k}+\vec{h}$ in the original TMS sampler. The output probabilities of the TMS sampler $p_{\text{tms}} (\vec{x})$   can then be calculated from the probabilities of the analogue SMS  Gaussian boson sampler $p_{\text{tms}}(\vec{h},\vec{k})$ as

\begin{equation}
p_{\text{tms}}(\vec{x})=\sum_{\substack{\vec{h},\vec{k}: \\\vec{x}=\vec{h}+\vec{k},\\ \sum_i h_i=\sum_i k_i=n/2}} p_{\text{gbs}}(\vec{h},\vec{k}),
\end{equation}
where $p_{\text{gbs}}(\vec{h},\vec{k})$ is the calculated probability of obtaining patterns $\vec{h}$ and $\vec{k}$ in the top and bottom $m$ output modes of the scheme in Fig.~\ref{SIFig_TMSGBS}b. The probability $p_{\text{gbs}}(\vec{h},\vec{k})$ is obtained using the total transfer matrix $T_{\ttt{gbs}} = (T \oplus T) \cdot T_{\ttt{bs}} \cdot T_{\ttt{ps}}$ which includes the matrices associated to the array of phase shifters $T_{\ttt{ps}}=U_{\ttt{ps}}^{\oplus m}$ and beam-splitters $T_{\ttt{bs}}=U_{\ttt{bs}}^{\oplus m}$, where

\begin{equation}
U_{\ttt{ps}}= \begin{bmatrix}
  1 & 0  \\
  0 & i 
 \end{bmatrix}, \qquad 
U_{\ttt{bs}}= \begin{bmatrix}
  1 & 1  \\
  -1 & 1 
 \end{bmatrix}/\sqrt{2}. 
\end{equation}

We remark that, although sampling from TMS input states deviates from an ideal implementation of GBS, it still represents a task which is hard classically, as it can be mapped into an analogous GBS problem. Therefore, rather then a validation on the computational complexity of the sampler, the test performed should here be interpreted as supporting evidence for the correct experimental implementation of the protocol.\\

\begin{figure*}[h!]
  \centering
  \includegraphics[width=0.5 \textwidth]{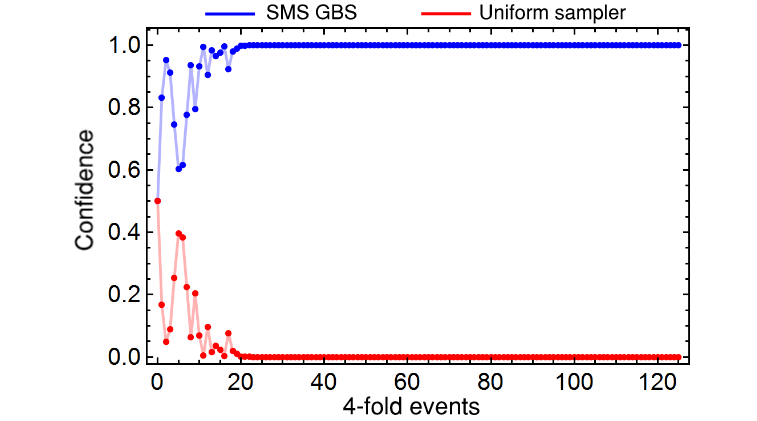}
  \caption{ \footnotesize	
Validation of GBS against a uniform sampler. The validation of the experiment against a sampler drawing from a uniform distribution is  performed using experimental collision-free data and Bayesian model comparison techniques.
} 
  \label{SIFig_GBSvsUnif}
\end{figure*}

{\bf Uniform distribution:} As for standard boson sampling, the output probabilities for a uniform sampler are simply given by $p_{\ttt{uni}}(\vec{k})=1/n_{\ttt{conf}}$ where $n_{\ttt{conf}}$ is the total number of possible output patterns such that $\sum_{i=1}^m k_i=n$. For completeness, validation results for GBS data against a uniform sampler are reported in Fig.~\ref{SIFig_GBSvsUnif}.

\begin{figure*}[t!]
  \centering
  \includegraphics[width=0.9 \textwidth]{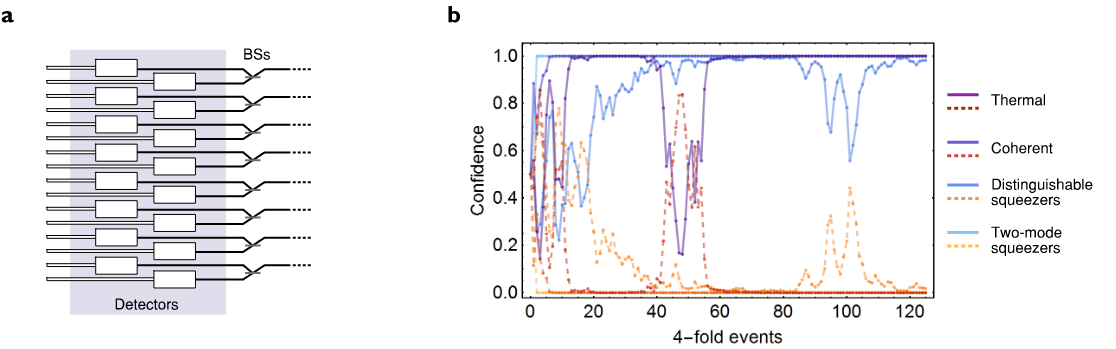}
  \caption{ \footnotesize	
Detection schematic for pseudo photon number resolving measurements and GBS validation. (a) Pseudo number resolving is obtained by probabilistically splitting the incoming photons via on-fiber $1\times2$ beam splitters and using SNSPDs and the outputs. (b) Validation of GBS using the data measured with pseudo photon number detection and Bayesian validation methods. Blue-purple curves are associated to the ideal GBS model, while red-yellow dashed curves to test models. In the order from darker to lighter colours, the curves represent validation against test models with input thermal, coherent, distinguishable SMS and TMS states.
} 
  \label{SIFig_PsuedoPNR}
\end{figure*}

\section{Gaussian boson sampling with pseudo photon number resolving detection}

To study GBS in a non-collision-free regime (i.e. including cases where more than one photon per output mode is possible), we additionally implemented GBS in a configuration where pseudo number resolving photodection was performed, where up to 2 photons could be probabilistically resolved in the output modes. This can be achieved with a slight modification at the detection apparatus. In particular, inserting 50:50 fiber beam splitters on the output fibers and detecting the outputs with the SNSPDs, as pictured in Fig.~\ref{SIFig_PsuedoPNR}a. In this way, if two photons are present in a single output mode, they can be probabilistically split at the beam splitter with a probability 1/2, and the simultaneous detection of both the associated detectors identifies the presence of two photons on the mode. The cases where a single photon is on a mode are not affected, as it will be detected by either one of the two SNSPDs associated to his mode. 

Results for the validation tests performed in the non-collision-free regime are reported in Fig.~\ref{SIFig_PsuedoPNR}b. As for the collision-free case reported in the main text, the validation are performed against hypothetical models of the experiment where thermal states, coherent states, distinguishable SMS and TMS states are used at the input instead of ideal SMS states, using Bayesian model comparison techniques. In all tests, we obtain high confidence in the ideal model after 125 events.

\section{Noises in Gaussian boson sampling with integrated SFWM sources}

We briefly discuss here some noise effects that may arise when performing GBS in an integrated implementation based on SFWM waveguide or ring-resonator sources~\cite{Silverstone2013,silverstone2015}. 

The first type of noise we study is the one inserted from losses before the circuit. For example,  losses due to cross-two-photon absorption, in addition to transmission losses, are known to limit the heralding efficiency of integrated non-linear sources~\cite{husko2013}. While in the standard approach to boson sampling, where single-photon sources are used and the inputs are Fock photon number states,  such losses can be simply accounted into a reduced efficiency of the sources~\cite{Loredo2017,wang2017,BSloss}, in GBS the situation is different: losses change the input Gaussian state. In fact, under the action of losses a single-mode squeezed state transforms towards a thermal state, inserting noises in the output distribution. Note that such effect is present also in SBS implementations based on parametric non-linear sources~\cite{SpagnoloValidation,Carolan2014,Bentivegna2015,Carolan2015}, where it is typically referred to as multi-pair emission noise. The GBS formalism allows to describe and study it analytically.

\begin{figure*}[t!]
  \centering
  \includegraphics[width=1\textwidth]{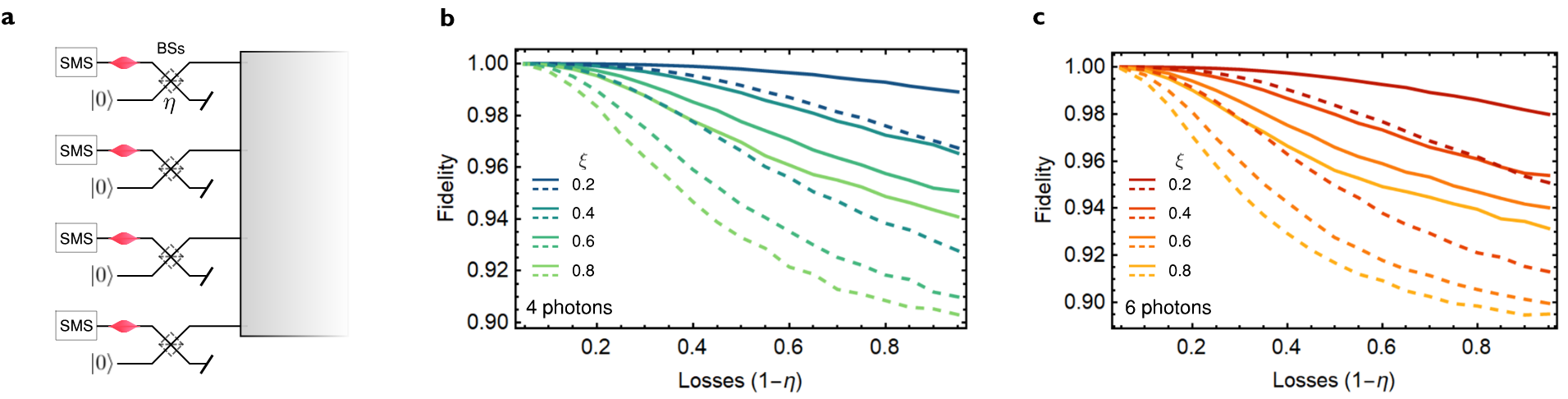}
  \caption{ \footnotesize	
Effect of losses before the interferometer on the output GBS distribution. (a) Simulation of losses at the input via the use of ancillary vacuum modes coupled to the signal modes via beam-splitters. The transmittivity of the beam-splitters $\eta$ controls the amount of losses ($1-\eta$) simulated. The Gaussian state injected into the interferometer is described by the covariance matrix of the state after the beam-splitters tracing out the row and columns associated to the ancillary modes, and output statistics are calculated using GBS techniques. (b) and (c), Simulated mean statistical fidelities of the distributions arising from lossy implementations with the ideal distribution obtained from an ideal device. The distributions for four photons (b) and six photons (c) interfering in a 12 mode interferometer are considered for different values of squeezing. Solid lines are for the case where 4 sources are used, while for the dashed lines 12 sources are used.   
} 
  \label{SIFig_LossesNoise}
\end{figure*}

Considering a GBS experiment where $n$ photons are emitted by $k$ SMS sources and injected into the interferometer, the presence of losses before the interferometer can be described as shown in Fig~\ref{SIFig_LossesNoise}a. Losses are modelled by combining each of the $k$ sources with an ancillary vacuum mode through a beam-splitter with transmission $0\leq\eta\leq 1$, and tracing out the (undetected) ancillary modes before sending the signal ones into the interferometer. Considering for simplicity uniform squeezing and losses through the modes, this procedure can be described analytically in the following way. First, we label the odd modes as the signal ones, i.e. with SMS sources on them, and the even ones as the auxiliary modes initialized in the vacuum state, so that the $2k$ squeezing parameters are given by $\vec{\xi}'=\{\xi_1,0,\xi_2,0,\ldots ,\xi_{k-1},0,\xi_k,0\}$. Following Eq.~(\ref{dss1}) and Eq.~(\ref{dss2}), the covariance matrix of the state after the action of the losses, modelled by the beam-splitters with transmission $\eta$, is given by:

\begin{eqnarray}\label{LossCov1}
\sigma^{(2k)}=\frac{1}{2}\begin{bmatrix}
	U_{\ttt{bs}} &  0  \\
	0         & U_{\ttt{bs}}^*
\end{bmatrix}S_{2k} S_{2k}^\dagger\begin{bmatrix}
	U_{\ttt{bs}}^\dagger &  0  \\
	0         & U_{\ttt{bs}}^\ttt{T}
\end{bmatrix}
\end{eqnarray}
where 
\begin{eqnarray}\label{LossCov2}
{S}_{2k}=\begin{bmatrix}
	\oplus_{j=1}^{2k} \cosh \xi'_j   &      \oplus_{j=1}^{2k} \sinh \xi'_j   \\
	\oplus_{j=1}^{2k}  \sinh \xi'_j   &      \oplus_{j=1}^{2k} \cosh \xi'_j
\end{bmatrix},\qquad
U_{\ttt{bs}}=
\begin{bmatrix}
	\sqrt{\eta}   &  \sqrt{1-\eta}   \\
	-\sqrt{1-\eta}   &  \sqrt{\eta}
\end{bmatrix}^{\oplus 2k}.
\end{eqnarray}

The Gaussian state injected in the interferometer is obtained by tracing out the ancillary modes. Its covariance matrix $\sigma$ is therefore simply given by deleting the even rows and columns from $\sigma^{(2k)}$ in Eq.~(\ref{LossCov1}). The output probabilities can finally be calculated via the Hafnian of the matrices obtained as described in Ref.~\cite{GBS_Main,GBS_Details} and in previous supplementary sections. 

In Fig.~\ref{SIFig_LossesNoise}b-c the mean statistical fidelities of the output 4-photon and 6-photon distributions for a 12 modes interferometer are reported as a function of the losses, for various values of squeezing and number of sources used. For a systems of this size, the protocol appears to have a good resilience to this type of noise. As expected, the noise becomes more significant increasing the system size (number of photons and sources) and the squeezing.\\

\begin{figure*}[ht!]
  \centering
  \includegraphics[width=0.5 \textwidth]{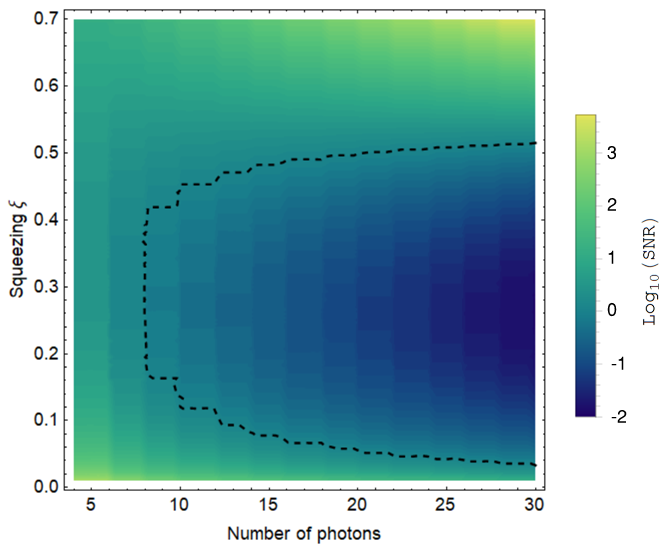}
  \caption{ \footnotesize	
Signal-to-noise ratio for GBS events performed with SFWM.  
The  SNR, calculated via Eq.~(\ref{SNREq}), is plotted in a logarithmic scale as a function of total photon number and squeezing values. The $\text{SNR}=1$ ($\text{Log}_{10} (\text{SNR})=0$) contour is reported  via the black dashed line. Number of sources is fixed to $k=2N$, with $N$ the number of photon pairs ($2N$ total photons), which saturates the complexity for GBS.}
  \label{SIFig_SNR}
\end{figure*}

The second type of noise we discuss is the noise due two spurious emission of photons via non-degenerate SFWM. This is relevant in our scheme based on SFWM as, using a dual-wavelength pump scheme in order to obtain SMS via degenerate SFWM, non-degenerate SFWM emission can be obtained at each sources by both pumps at different wavelengths. Effectively, each source comprises a single-mode squeezer and also two two-mode squeezers, one per pump. The presence of TMS induces noise in the system. For example, consider the case when we observe two photons in the signal modes (were only the signal wavelength is selected). In the ideal case these would be generated from the first term of a SMS state, but there is also a chance that these two photons could incorrectly originate from two-pairs of non-degenerate pairs emitted via two-mode squeezing, where only the two photons in the signal wavelengths are observed while the other two photons are discarded. Such event would then give an erroneous sample. 

The signal-to-noise ratio (SNR) between the correct events originated by ideal SMS and the noisy events where at least a photon pair is originated via spurious TMS can be quantified. For the ideal case, the probability of generating $n$ pairs of photons ($2n$ total signal photons) from $k$ SMS sources with squeezing $\xi$ is given by the negative binomial distribution (see for example Ref.~\cite{GBS_Main,GBS_Details}):

\begin{equation}
p_\ttt{sms}^{(k,\xi)} (2n) = {{k/2 + n -1 }\choose{n}} \sech^k (\xi) \tanh^{2n}(\xi).
\end{equation}

The probability of obtaining $n$ signal photons ($n$ pairs) via $k$ TMS sources with squeezing $\xi$ is instead given by (see for example Ref.~\cite{GBS_Main,GBS_Details}):

\begin{equation}
p_\ttt{tms}^{(k,\xi)} (n) = {{k}\choose{n}} \sech^{2k} (\xi) \tanh^{2n}(\xi).
\end{equation}

Suppose now performing GBS with $k$ sources based on SFWM, that is $k$ SMS sources and $2k$ TMS sources with same squeezing $\xi$, and suppose we observe a $2n$ photons event. We consider it a correct event if and only if all photons originate from SMS, while all other cases are considered as noise. The probability of observing a noisy event is then given by:

\begin{align}
p_{\times}^{(k,\xi)} (2n) &= \sum_{m=1}^{n} p_\ttt{sms}^{(k,\xi)} (2(n-m)) p_\ttt{tms}^{(2k,\xi)} (2m) \\
&= \sech^{5k}(\xi) \tanh^{2n}(\xi) \sum_{m=1}^{n} {{k/2 +n -m -1}\choose{n-m}}{{2k}\choose{2m}} \tanh^{2m}(\xi).
\end{align}

\noindent The SNR is then finally given by:

\begin{align}
\text{SNR}(2n,k,\xi)&=\frac{p_\ttt{sms}^{(k,\xi)} (2n)}{p_{\times}^{(k,\xi)} (2n)}\\
&=\cosh^{4k}(\xi)\frac{{{k/2 +n -1}\choose{n}}}{\sum_{m=1}^{n} {{k/2 +n -m -1}\choose{n-m}}{{2k}\choose{2m}} \tanh^{2m}(\xi)}. \label{SNREq}
\end{align}

The SNR as a function of photons and squeezing is plotted in Fig.~\ref{SIFig_SNR}, where we take $k=2n$ (which saturates the complexity for GBS~\cite{GBS_Main}). It can be observed that for low levels of squeezing the SNR decreases rapidly with the number of photons. However, for higher squeezing the $\cosh^{4k}(\xi)$ factor in Eq.~(\ref{SNREq}) becomes dominant and a good SNR is recovered.

To reduce this type of noise, different methods can be adopted. First, one may consider to detect the idler photons emitted in the TMS case, instead of discarding them, via filtering them before the interferometer and detecting the idler modes, in the same way as in the scattershot approach. If any idler photon is detected in addition to the photons in the signal modes, then the event is discarded. In this way the SNR can be in principle arbitrarily high (though in practice it would be limited by the loss).  One other option would be to consider an hybrid scheme between GBS and SBS, where detecting the idler photons would herald the injection of a non-Gaussian, but known, state into the interferometer, while if no idler is detected then the event is treated in the standard GBS approach. Such scheme would be more suitable for integrated approaches based on SFWM, increasing the count rate by not erasing all the unideal events, and is likely to be of the same complexity of GBS. 

\section{Franck-Condon profiles}

The analogy between photons distributed among optical modes and molecular phonons among vibrational modes comprises a platform for photonic quantum simulations of complex molecular dynamics~\cite{AspuruGuzik,ClementsVibronic,Sparrow2018}. As we detail in this section, based on such an analogy, a setting similar to Gaussian boson sampling can be used, e.g., to simulate an arbitrary molecular Franck-Condon (FC) profile~\cite{AspuruGuzik}. The latter corresponds to the molecular vibrational transition profile within the harmonic approximation and the assumption of a coordinate-independent electronic transition moment.

According to the Franck-Condon principle~\cite{franck,condon}, the probability of a vibrational transition from an initial state $|\vec{j}\rangle$, which we assume to be the ground state $|\vec{j}\rangle=\ket{0,\dots,0}$, to a final state $|\vec{k}\rangle=\ket{k_1, \dots, k_m}$ is given as

\begin{equation}\label{FC1}
p_\ttt{FC}(\vec{k})=|\langle\vec{k}|\mathcal{U}_\ttt{Dok}|{0, \dots, 0}\rangle|^2
\end{equation}
where $m$ is the number of vibrational modes and $\mathcal{U}_\ttt{Dok}$ is the Doktorov transformation~\cite{doktorov}. The latter consists of linear transformations that conserve the number of excitations, mode displacement and single-mode squeezing,
\begin{equation}\label{FC2}
\mathcal{U}_\ttt{Dok}=\mathcal{U}_\mathrm{L}\left[\otimes_{j=1}^m\mathcal{S}(\xi_j)\right]\mathcal{U}_\mathrm{R}^\dagger \left[\otimes_{i=1}^m\mathcal{D}(\alpha_i)\right].
\end{equation}
Here, $\mathcal{D}(\alpha_{i})=\exp\left(\alpha_{i} {a}_i+\alpha_{i}^* {a}_i^\dagger\right)$ and $\mathcal{S}(\xi_i)=\exp\left[\xi_i/2 \left({a}_i^2+{a}_i^{\dagger^2}\right)\right]$ denote, respectively, the displacement and squeezing operators acting on the $i$th vibrational mode ($\alpha_{i}$ and $\xi_i$ are displacement and squeezing parameters), $\mathcal{U}_\mathrm{L}$ and $\mathcal{U}_\mathrm{R}$ are linear transformations and ${a}_i^{\dagger}$ $({a}_i)$ are the bosonic creation (annihilation) operators. 

The displacement and squeezing parameters, as well as linear transformations $\mathcal{U}_\mathrm{L}$ and $\mathcal{U}_\mathrm{R}$ are defined via the initial and final molecular configurations. Namely, these parameters can be found from the following expressions
\begin{eqnarray}\label{FC3}
&& \alpha_i=\frac{1}{\sqrt{2}}\sum_{j=1}^m J^{-1}_{ij}\delta_j, \,\,\,\,\,\,\, J=\Omega'U_\ttt{D} \Omega^{-1},  \,\,\,\,\,\,\,  \delta_i=\sum_{j=1}^m\Omega'_{ij} d_j\\
&& \Omega'=\ttt{diag}(\sqrt{\omega_1'}, \dots, \sqrt{\omega_m'}), \,\,\,\,\,\,\, \Omega=\ttt{diag}(\sqrt{\omega_1}, \dots, \sqrt{\omega_m}).
\end{eqnarray}
Here, $\omega_i$ and $\omega'_i$ are the harmonic angular frequencies of the initial and final electronic states. In turn, $U_\ttt{D}$ and $d_i$ are, respectively, the Duschinsky rotation matrix and displacements along the normal coordinates $q_i$ (associated with mode operator ${a}_i$), which are responsible for molecular structural changes induced by an electronic transition. These structural changes or, in other words, normal coordinate transformations can be written as~\cite{duschysnky}
\begin{equation}\label{FC4}
q_i\rightarrow \sum_{j=1}^m U_{\ttt{D}}^{[ij]} q_j+d_i,
\end{equation}
where $U_{\ttt{D}}^{[ij]}$ denote matrix elements of $U_{\ttt{D}}$. 

In turn, squeezing parameters $\xi_i$ and linear transformations $\mathcal{U}_\mathrm{L}$ and $\mathcal{U}_\mathrm{R}$ are obtained from the singular value decomposition of the matrix $J$,
\begin{equation}\label{FC5}
J={U}_\mathrm{L} \Sigma  {U}_\mathrm{R}^\ttt{T}.
\end{equation}
That is, $\mathcal{U}_\mathrm{L}$ and $\mathcal{U}_\mathrm{R}^\dagger$ are the state-space representations of $U_\mathrm{L}$ and $U_\mathrm{R}^\ttt{T}$ (T denotes matrix transposition), while squeezing parameters $\xi_i$ are the natural logarithms of the diagonal elements of the matrix $\Sigma$, i.e., of the singular eigenvalues of $J$.

Having defined all the quantities appearing in Eq.~(\ref{FC1}), we can now interpret $p_\ttt{FC}(\vec{k})$ (also known as an FC factor) as the probability of detecting an excitation pattern $\vec{k}$ at the output of a linearly evolved squeezed coherent state $\ket{\Psi}:=\left[\otimes_{j=1}^m\mathcal{S}(\xi_j)\right]\mathcal{U}_\mathrm{R}^\dagger \left[\otimes_{i=1}^m\mathcal{D}(\alpha_i)\right]\ket{0, \dots, 0}$, i.e.,
\begin{equation}\label{FC6}
p_\ttt{FC}(\vec{k})=\left|\langle{\vec{k}}|\mathcal{U}_\ttt{L}|\Psi^{}\rangle\right|^2. 
\end{equation}
Finally, the molecular FC profile at a given frequency $\omega$ reads
\begin{equation}\label{FC7}
\ttt{FCP}(\omega)=\sum_{k_1, \dots, k_m=0}^\infty p_\ttt{FC}(\vec{k})\delta_{\omega-\sum_{i=1}^m\omega'_i k_i,0},
\end{equation}
where $\delta_{x,y}$ is the Kronecker delta. Importantly, $\ttt{FCP}(\omega)$ can be approximated by generating samples $\vec{k}$ according to the probability distribution $p_\ttt{FC}(\vec{k})$. On the other hand, knowing that FC factors $p_\ttt{FC}(\vec{k})$ are the probabilities of detecting excitation patterns $\vec{k}$ at the output of linearly evolved squeezed coherent states, we conclude that FC profiles can be simulated in a photonic experiment. That is, one can approximate $\ttt{FCP}(\omega)$ by generating photonic displaced squeezed states, evolving them through a linear-optical network and performing boson sampling at its output~\cite{AspuruGuzik}. Such a simulation procedure is, therefore, similar to Gaussian boson sampling. However, as opposed to Gaussian boson sampling, photonic simulations of FC profiles require data corresponding to the detection of various photon numbers $\sum_{i=1}^m k_i=n$, i.e., $n$ is not fixed here. Moreover, in general, simulations of FC profiles necessitate displaced squeezed states rather than vacuum squeezed states as per Gaussian boson sampling (Note, however, that molecular structural changes with no displacement do exist. For instance, photonic FC profile simulations of the tropolone molecule, C$_7$H$_6$O$_2$, require squeezed vacuum states only~\cite{ClementsVibronic}).

As explained in the main text, the described procedure for simulating FC profiles via a photonic quantum device can be reversed to provide a benchmarking tool. Namely, for a given photonic experiment, an FC profile corresponding to a synthetic molecule can be constructed and simulated. The merit of such a procedure is to understand the fidelities that one could expect to find in a similar setting engineered to investigate a particular molecule. To achieve this, given the description of our experiment, we deduce the corresponding molecular parameters from Eqs.~(\ref{FC1})-(\ref{FC5}) (this procedure yields not a unique but a set of synthetic molecules). Since our experiment relies on squeezed vacuum states, $\ket{\Psi}$ appearing in Eq.~(\ref{FC6}) is a squeezed vacuum state as well, $\ket{\Psi}=\otimes_{i=1}^m\mathcal{S}(\xi_i)\ket{0, \dots, 0}$. Thus, we are able to simulate squeezing molecular structural transformations only, i.e., $\alpha_i=0$, and, consequently, $d_i=0$, $\forall i$, in Eq.~(\ref{FC4}). In this configuration, the output probabilities can be described using the GBS formalism [Eq.~(\ref{a1})-(\ref{a4})], so that the FC factors $p_\text{FC}(\vec{k})$ are given by

\begin{equation}\label{FC8}
p_\text{FC}(\vec{k})=\frac{|\ttt{Haf }{B}_{\vec{k}}|^2}{k_1!\cdots k_m!\sqrt{\ttt{det}\sigma_{\ttt{Q}}}},
\end{equation}
where $B_{\vec{k}}$ is obtained from ${B}={\mathcal{U}_\ttt{L}}\cdot \ttt{diag}[\tanh\xi_1, \dots, \tanh \xi_m]\cdot\mathcal{U}^\ttt{T}_\ttt{L}$ by repeating $k_i$ times its $i$th column and row, and $\sigma_{\ttt{Q}}$ is given in Eq.~(\ref{a1}). 

To quantify the enhancement of a quantum experiment over classical approximation strategies we adopt the approach developed in Ref.~\cite{ClementsVibronic}. The idea is to quantify this improvement in terms of the fidelity of experimentally reconstructed profiles with respect to the fidelity obtainable by the best classical approximation. The latter represents FC profiles with highest fidelity achievable from classical experiments, i.e., experiments employing classical states only (that is, states with a non-negative Glauber-Sudarshan function~\cite{ulf}). As the fidelity between states is invariant under unitary evolutions, finding the optimal classical strategy corresponds to finding the closest classical state to initial displaced squeezed states~\cite{ClementsVibronic}. In turn, the single-mode classical state closest to a single-mode displaced Gaussian state is a coherent state with the same level of displacement~\cite{Marian2002}. Consequently, in absence of displacement, vacuum is the optimal classical state. To quantify the quantum enhancement obtained in the experiment we then consider the difference $\mathcal{C}= F_\ttt{Q}-F_\ttt{C}$ between the fidelity $F_\ttt{Q}$ of the FC profile reconstructed from the experiment to the theoretical FC profile, and the fidelity obtained using the optimal classical strategy $F_\ttt{C}$.\\

As a first test, we reconstruct the FC profile for a synthetic molecule directly associated to the device. Collected experimental data using a pseudo number resolving detection scheme were used. In this case, reversing the protocol and using it as a benchmarking tool corresponds to using the transfer matrix of the device as $\mathcal{U}_\ttt{L}$ and the characterised squeezing values as $\xi_i$. As the choice of frequencies $\omega'_i$ does not affect the fidelity of an FC profile (only its shape), which is the parameter of interest here, we choose them randomly and with arbitrary unit of measure. This defines the synthetic molecule used for benchmarking. In this case, although the fidelity of the profile is very high ($F_\ttt{Q}>99\%$), the quantum enhancement is only $\mathcal{C}=0.4\%$, i.e. while there is still an improvement, it is actually very low. This is due to the fact that the squeezing in the sources is very low, which means that the FC profile is dominated by classical vacuum contributions.  

To obtain higher quantum enhancements, molecules involving higher squeezing values have to be involved. Below we describe a prescription to simulate such molecules even when the actual squeezing in the device is lower. This may be useful in practical implementations of Franck-Condon quantum simulations. \\

For simplicity, we focus on the case where no displacement is used, although the approach can be easily generalised to the case with non-zero displacement. Consider the situation where we want to reconstruct the FC profile $p_\text{FC}(\vec{k})$ for a molecule with  corresponding squeezing parameters $\xi_i$ and unitary operation $\mathcal{U}_\ttt{L}$. However, in our imperfect experiment we have access to a limited amount of squeezing only, $\bar{\xi}_\ttt{max}\leq \max_i \xi_i$, so that for each source we can only implement a squeezing $\bar{\xi}_i \leq \bar{\xi}_\ttt{max}$. The transition amplitudes can be then estimated in the imperfect device by tuning the squeezing parameters in the sources so that $\tanh(\bar{\xi}_i)=\gamma  \tanh(\xi_i)$, with $0<\gamma \leq 1$ a real constant rescaling and $\ttt{diag}[\tanh(\bar{\xi}_1), \dots, \tanh (\bar{\xi}_m)]=\gamma \ttt{\ diag}[\tanh(\xi_1), \dots, \tanh (\xi_m)]$. In this way, considering also uniform losses $\eta$ in the device, from Eq.~(\ref{FC8}), we have a simple relation between the ideal FC factors $p_\text{FC}(\vec{k})$ and the ones reconstructed in the imperfect device $\bar{p}_\text{FC}(\vec{k})$, given by

\begin{equation}\label{FC9}
\bar{p}_\text{FC}(\vec{k})=\mathcal{N}\eta^n \gamma^{2n} p_\text{FC}(\vec{k}),
\end{equation}
with $n=\sum_{i=1}^{m} k_i$ the total number of photons in the output configuration $\vec{k}$, and $\mathcal{N}$ is a normalization constant. As $\eta$ and $\gamma$ are known from the characterisation of the device, Eq.~(\ref{FC9}) can be inverted to reconstruct the FC profile via 
\begin{equation}
p_\text{FC}(\vec{k})=\frac{\bar{p}_\text{FC}(\vec{k})/\eta^{\sum k_i} \gamma^{2\sum k_i}}{\sum_{\vec{x}} \bar{p}_\text{FC}(\vec{x})/\eta^{\sum x_i} \gamma^{2\sum x_i} }.
\end{equation}

Note that this post-processing procedure does not increase the computational complexity of the experiment, in the sense that, if from the experiment we are able to collect only up to $n$-photon events, then we are truncating the FC profile in Eq.~(\ref{FC7}) at $n$-photon contributions. This implies that an FC profile reconstructed with this approach is faithful only if the ideal contributions involving more than $n$ photons are negligible, which, for large values of $n$, is in general true only for $\eta$ and $\gamma$ close to unity. This can be also observed in Fig.~\ref{FigFC} (inset), where the quantum enhancement initially increases with $\gamma$, but starts to decrease as higher order terms become dominant. Therefore, the approach is clearly not scalable, i.e. it will not provide much help in future large-scale implementations. Nonetheless, it can provide significant advantages in near-term quantum devices, as shown in its application to our experiment.\\

We implemented the protocol described above to reconstruct FC profiles of synthetic molecules requiring higher squeezing using the post-processing procedure. The synthetic molecule to be simulated is obtained as in the first test above, with the only difference that now the associated squeezing values are $\xi_i=\bar{\xi}_i/\gamma$, where $\bar{\xi}_i$ are the squeezing parameters in the device, preliminarily characterised as described in previous sections. In Fig.~\ref{FigFC} (inset) the improvement over optimal classical strategies is reported for different values of simulated squeezing (reported as $\max_i \xi_i$), that is for different values of $\gamma$, both for the case where the post-processing is performed (solid black line) and when raw data is instead applied (dashed black line). It can be noted that the procedure provides a significant improvement, with the quantum enhancement initially increasing and then decreasing when the FC terms, due to more than 4-photon contributions, start to be dominant. The maximum improvement $\mathcal{C}= 9\%$ is obtained for $\gamma \approx 5$, for which the FC profile (plotted in Fig.~\ref{FigFC}) has a fidelity of $86\%$ with the ideal one. In all the cases, applying the post-processing procedure provides a significant advantage as compared to raw data usage.\\

\section{Estimation of photon number scaling with current silicon photonics technology}

Finally, we analyse the potential of the integrated quantum photonics approach as a platform for multi-photon boson sampling protocols in terms of estimated count rates when scaling up the component integration with current technologies. By using as relevant parameters for each component the values measured in our experimental set-up, we aim to provide an approximate evaluation of what event rates can be expected from current photonic technologies in near-term devices. The rates for events with $n$ signal photons, emitted from $k$ integrated sources and injected in a $m$ mode interferometer, for SBS and GBS are respectively

\begin{align}
R_\ttt{sbs}(n,k,m)&=R_0 \left[ {k \choose n} \tanh(\xi)^{2n} \sech(\xi)^{2k}  \right] \eta_\ttt{u}^{mn} \eta_\ttt{ch}^{2n} \eta_\ttt{det}^{2n},   \nonumber \\
R_\ttt{gbs}(n,k,m)&=R_0 \left[ {k/2+n/2 -1  \choose n/2} \tanh(\xi)^{n} \sech(\xi)^{k}  \right] \eta_\ttt{u}^{mn} \eta_\ttt{ch}^{n} \eta_\ttt{det}^{n}.   \label{eq:Rates}
\end{align}	
Here $R_0$ represents the repetition rate of the pump, $\xi$ the squeezing at the sources, $\eta_\ttt{det}$ the detection efficiency, $\eta_\ttt{ch}$ the transmission of the optical channels between the interferometer and the detectors (which includes chip-to-fiber coupling loss for off-chip detection). The factor $\eta_\ttt{u}^m$ represents the total losses in the interferometer, where $\eta_\ttt{u}$ represents the losses in each coupling operation (e.g. evanescently coupled waveguides in a universal scheme~\cite{Reck1994,Carolan2015,Clements2016}), and the number of single operations each photon undergoes in the $m$ mode interferometer is assumed to be $m$, as in the universal scheme proposed in Ref.~\cite{Clements2016}.

In the analysis below we consider a source in each input mode of the interferometer ($m=k$), and used the system efficiencies characterised in our silicon photonic device and experimental set-up: $\eta_\ttt{det}=80\%$, $\eta_\ttt{ch}=64\%$, $\eta_\ttt{u}=99.95\%$, $R_0= 500 \text{ MHz}$. We consider two standard types of integrated sources: a low efficiency one, i.e. the spiral sources used in our device, with a squeezing parameter $\xi=0.17$, and more efficient sources based on integrated ring resonator cavities (see, for example, Ref.~\cite{silverstone2015}), with $\xi=0.31$. We also consider the case where on-chip detection is performed, where no losses associated to off-chip coupling and fiber transmission are present ($\eta_\ttt{ch}=1$). 

\begin{table}[t]
  \centering
  \renewcommand{\arraystretch}{1.2}
  \begin{tabular}{|p{2.2cm}||p{1.6cm}|p{1.6cm}|p{1.6cm}|p{1.6cm}|p{1.6cm}|p{1.6cm}|p{1.6cm}|p{1.6cm}|}
    \hline
    \multirow{3}{2.2cm}{Number of signal photons} & \multicolumn{8}{c|}{Event Rate (Hz)} \\
    \cline{2-9}
    & \multicolumn{2}{c|}{Spiral Sources}& \multicolumn{2}{c|}{Spiral Sources \& Int.Det.} & \multicolumn{2}{c|}{Ring Sources} & \multicolumn{2}{c|}{Ring Sources \& Int.Det.}\\    
    \cline{2-9}
    & \centering{SBS} & \centering{GBS} & \centering{SBS} & \centering{GBS}& \centering{SBS} & \centering{GBS}& \centering{SBS} & \ \ \ \ \ GBS\\
    \hhline{|=||=|=|=|=|=|=|=|=|}
 10 &  $0.3$ & $6\times 10^3$ & $2\times 10^3$ & $5\times 10^5$ & 65 & $6 \times 10^4$ & $4 \times 10^5$  & $5 \times 10^6$\\ \hline
 
  16 & -  & 1 & $8\times 10^{-3}$ & $1\times 10^3$ & $2\times 10^{-3}$ & $4 \times 10^{2}$ & $3\times 10^{3}$ & $4 \times 10^5$\\ \hline
 
 20 & - & $2\times 10^{-3}$ &  - & $15$ & - & $7$ & 30 & $5 \times 10^4$\\ \hline 
 
 32 & - & - &  - & $4\times 10^{-6}$ & - & $1\times 10^{-5}$ & - & 15\\ \hline  
 
 40 & - & - &  - & - & - & - & - & $2\times 10^{-2}$\\ \hline  
  \end{tabular}
  \caption{Estimated count rates near-term experiments using current technologies. The rates are calculated considering a silicon device where integrated sources deliver quantum states to a 100-modes interferometer ($m=k=100$). This configuration requires in total 200 detectors for SBS (100 for the signal modes and 100 for the idler modes), and 100 detectors for GBS. Note that in SBS $n$ signal photons detection correspond to $2n$-fold events including the $n$ idler photons. The calculation, via Eq.(\ref{eq:Rates}), is performed using the average values for coupling, filtering, on-chip and off-chip transmission losses, and detection efficiency as characterised in our experiment.}
 \label{tab:rates}
\end{table}

\begin{figure*}[t]
  \centering
  \includegraphics[width=0.8 \textwidth]{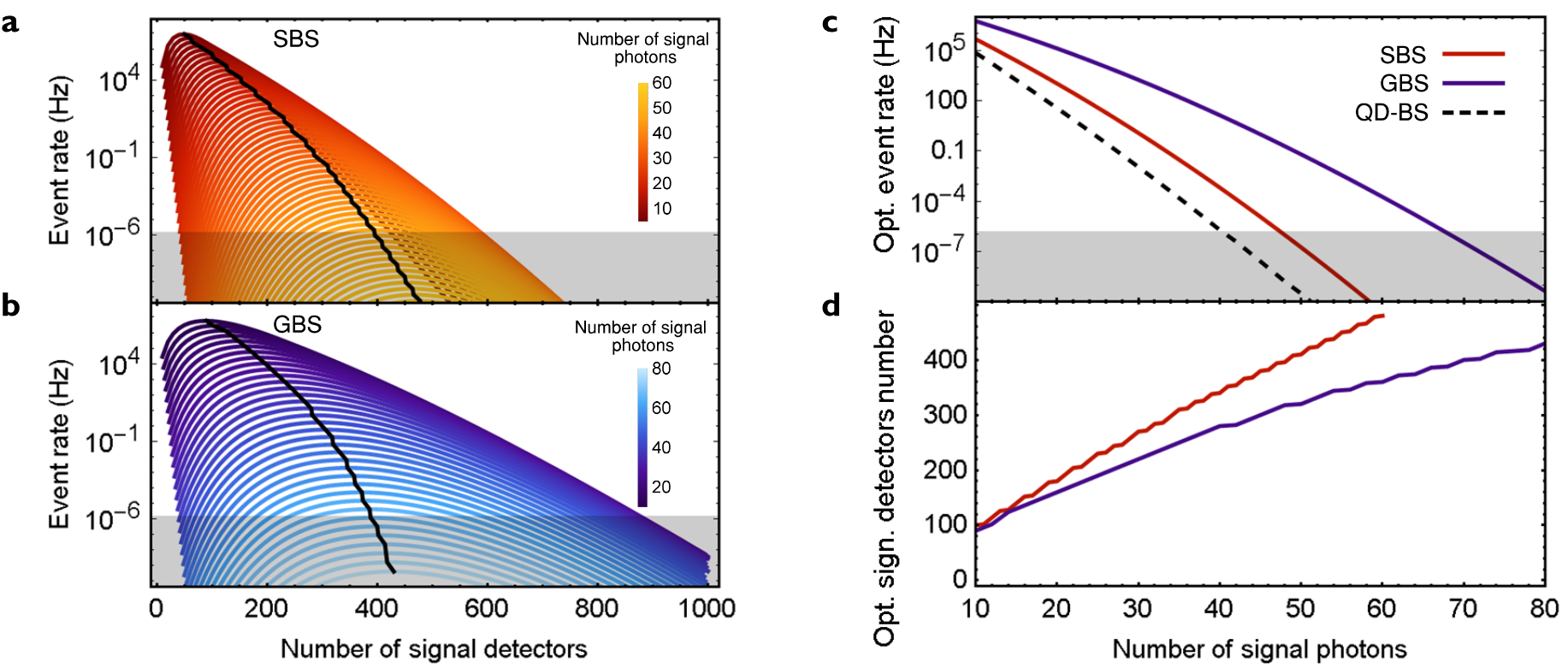}
  \caption{ \footnotesize	
Estimated counts for different sizes of the optical circuit. In this analysis, the number of signal detectors is considered to be the same as the number of modes in the interferometer and the number of sources. a) and b) show the event rate for different circuit size and number of photons in a SBS scenario and a GBS scenario, respectively. Black lines show the optimal values of the event rates. c) Optimal event rate estimated for different numbers of signal photons in the SBS (red) and GBS (blue) regimes. The dashed line represents an estimation for the case where standard boson sampling is performed using on-chip quantum dots sources (where, for fair comparison, the size of the interferometer is the same as in GBS). d) Size of the interferometer associated to the optimal cases. In all plots, shaded areas represent impractical experiments, where the threshold is set to be $1$ event/week.
}
  \label{SIFig_EstimateCounts}
\end{figure*}

In Table~\ref{tab:rates} we report the event rate estimates for these cases with a system size corresponding to a 100-modes interferometer ($k=m=100$), which is realistic on near-term devices thanks to the fabrication scalability of silicon photonics. In particular,  an array of $100$ detectors to detect the signal output modes (200 in total for SBS considering the idler modes) would be required, which is possible with current technologies. For example, arrays of $>200$ individually addressable on-chip high-efficiency SNSPDs have already been demonstrated (see e.g. Ref.~\cite{Schuck2013}). The rates suggest that experiments with tens of interfering photons should be available with further scaling of current components in silicon photonics, without any further major technological breakthrough required. 

To investigate the limits of the approach, namely scaling up the number of components with current silicon quantum photonics technology, we have tested the event rates consider larger-scale circuits with up to 1000-modes interferometers and 1000 detectors on the signal modes. Again, we assume the number of sources and signal modes and detectors to be the same, and consider a technology including ring sources and integrated detectors with efficiency as above. As can be observed in Fig.~\ref{SIFig_EstimateCounts}a-b, while increasing the number of sources provides an initial improvement to the event rates, it actually turns to be detrimental if the circuit size becomes too large. The reason for this is that losses in the interferometer, which scale as $\eta_\ttt{u}^m=\eta_\ttt{u}^k$ become dominant for values of $k$ that are too large, suppressing the combinatorial enhancement provided by the scattershot or Gaussian boson sampling approaches. A trade-off thus has to be adopted to obtain an optimal event rate in SBS and GBS, which is reported in Fig.~\ref{SIFig_EstimateCounts}d (also shown as black lines in Fig.~\ref{SIFig_EstimateCounts}a-b). The optimal event rates as a function of photon number are reported in Fig.~\ref{SIFig_EstimateCounts}c for the different approaches. As expected, the presence of losses implies an exponential decrease in the event rate. However, experiments with a large number of photons are still possible before the event rate becomes impractical, which we identify as experiments with an event rate lower than a threshold of 1 event/week (shaded areas in Fig.~\ref{SIFig_EstimateCounts}). For GBS such threshold is reached with $\approx 70$ signal photons, while for SBS with $\approx 48$ signal photons. These values are expected to be at the limit of what is tractable for classical supercomputers~\cite{Neville2017}. 

For comparison, we also perform the analysis for another approach which has been recently investigated, that is standard boson sampling using a time-demultiplexed high-efficiency quantum dot source to deliver multi-photon Fock states into the circuit~\cite{wang2017}. For the relevant parameters we use values similar to the ones proposed in Ref.~\cite{wang2017}: photon generation probability from the quantum dot $p_\ttt{qd}=65\%$ and repetition rate $R_{0,\ttt{qd}}=76\text{ MHz}$. While we optimistically consider no losses in delivering the photons to the integrated circuit (achievable for example by integrating the quantum dot), the losses in the photon demultiplexing scheme, given by $\eta_\ttt{demux}=\lceil \eta_\ttt{switch} \log_2 n \rceil^n$, where $\eta_\ttt{switch}$ is the value for each switch in the scheme. We consider low-loss switches based on Pockels cells with $\eta_\ttt{switch}=99.5\%$, as proposed in Ref.~\cite{wang2017}. The scaling of the event rate in this scenario is reported as a dashed line in Fig.~\ref{SIFig_EstimateCounts}c, where, for fair comparison, the size of the interferometer is considered to be the same as for the GBS case. We estimate a significant decrease of the event rates with this approach compared to SBS or GBS.

To further increase the complexity, e.g. to target experiments with hundreds of photons, the capability to scale up the number of current optical components in silicon quantum photonics is not enough: technological progress is required.  A first improvement would be to develop materials with lower transmission losses and more efficient sources of squeezed light.
Promising integrated photonic platforms are being developed in this direction~\cite{Dutt2015}. Ultimately, the introduction of rudimentary error-correction techniques in boson sampling, tackling the effect of losses, is likely to be required for applications far beyond current capabilities.

\end{document}